\documentclass[]{aa}  
\pdfoutput=1
\usepackage{graphicx}
\usepackage{caption}
\usepackage{subcaption}
\usepackage{epsf}
\usepackage{longtable,lscape}
\usepackage{dcolumn}
\usepackage{booktabs} 
\usepackage{url}
\usepackage{float}
\usepackage{microtype}
\usepackage[varg]{txfonts}
\usepackage{natbib,twoopt} 
\bibpunct{(}{)}{;}{a}{}{,} 

\let\oldhat\hat

\renewcommand{\hat}[1]{\oldhat{\mathbf{#1}}}
\newcommand{\kms}{km\,${\rm s}^{-1}$}
\newcommand{\smy}{$[M_\odot\,{\rm yr}^{-1}]$\,}

\begin{document}

   \title{Wolf-Rayet stars in the Small Magellanic Cloud}

   \subtitle{II. Analysis of the binaries}

   \author{T. Shenar\inst{1} 
          \and R.\ Hainich\inst{1}
          \and H.\ Todt\inst{1} 
          \and A.\ Sander\inst{1}
          \and W.-R.\ Hamann\inst{1}
          \and A.\ F.\ J.\ Moffat\inst{2}
          \and J.\ J.\ Eldridge\inst{3}
          \and H.\ Pablo\inst{2}          
          \and L.\ M.\ Oskinova\inst{1} 
          \and N.\ D.\ Richardson\inst{4}         
          }

   \institute{\inst{1}{Institut f\"ur Physik und Astronomie, Universit\"at Potsdam,
                Karl-Liebknecht-Str. 24/25, D-14476 Potsdam, Germany}\\
              \inst{2}{D\'epartement de physique and Centre de Recherche en Astrophysique 
                du Qu\'ebec (CRAQ), Universit\'e de Montr\'eal, C.P. 6128, Succ.~Centre-Ville, Montr\'eal, Qu\'ebec, H3C 3J7, Canada}\\            
              \inst{3}{Department of Physics, University of Auckland, Private Bag 92019, Auckland, New Zealand}\\
              \inst{4}{Ritter Observatory, Department of Physics and Astronomy, The University of Toledo, Toledo, OH 43606-3390, USA} \\
              \email{shtomer@astro.physik.uni-potsdam.de}   
              }
   \date{Received December 08, 2015 / Accepted March 30, 2016}


\abstract
{Massive Wolf-Rayet (WR) stars are evolved 
massive stars ($M_\text{i} \gtrsim 20\,M_\odot$) characterized by strong mass-loss. Hypothetically, they can form either as 
single stars or as mass donors in close binaries.
About 40\,\% of the known WR stars are confirmed binaries, raising 
the question as to the impact of binarity on the WR population. 
Studying WR binaries is crucial in this context, and furthermore enable one
to reliably derive the elusive masses of their components, 
making them indispensable for the study of massive stars.
}
{
By performing a spectral analysis of all multiple WR systems in the
  Small Magellanic Cloud (SMC), 
  we obtain the full set of stellar
  parameters for each individual component. Mass-luminosity relations are tested, and 
  the importance of the binary evolution channel is assessed.
}
{The spectral analysis is performed
with the Potsdam Wolf-Rayet (PoWR) model atmosphere code by superimposing model spectra that correspond to each component.
Evolutionary channels are constrained 
using the Binary Population and Spectral Synthesis  (BPASS) evolution tool.}
{
Significant Hydrogen mass fractions ($0.1 <X_\text{H} < 0.4$) are detected in all WN components. 
A comparison with mass-luminosity relations and evolutionary tracks implies that the majority of the WR stars in our sample 
are not chemically homogeneous.
The WR component in the binary AB\,6 is 
found to be very luminous ($\log L \approx 6.3\,[L_\odot]$) given its orbital mass ($\approx 10\,M_\odot$), presumably 
because of observational contamination by a third component. 
Evolutionary paths derived for our objects 
suggest that Roche lobe overflow had occurred in most systems, affecting their evolution.
However, the implied initial masses 
($\gtrsim 60\,M_\odot$) are large enough for the  primaries to have entered the WR phase, regardless of binary interaction.
}
{
Together with the results for the putatively single SMC WR stars, our study suggests that the binary 
evolution channel does not dominate the formation of WR stars at SMC metallicity.
}

\maketitle

\section{Introduction}
\label{sec:introduction}

\begin{figure*}[!htb]
\centering
  \includegraphics[width=\textwidth]{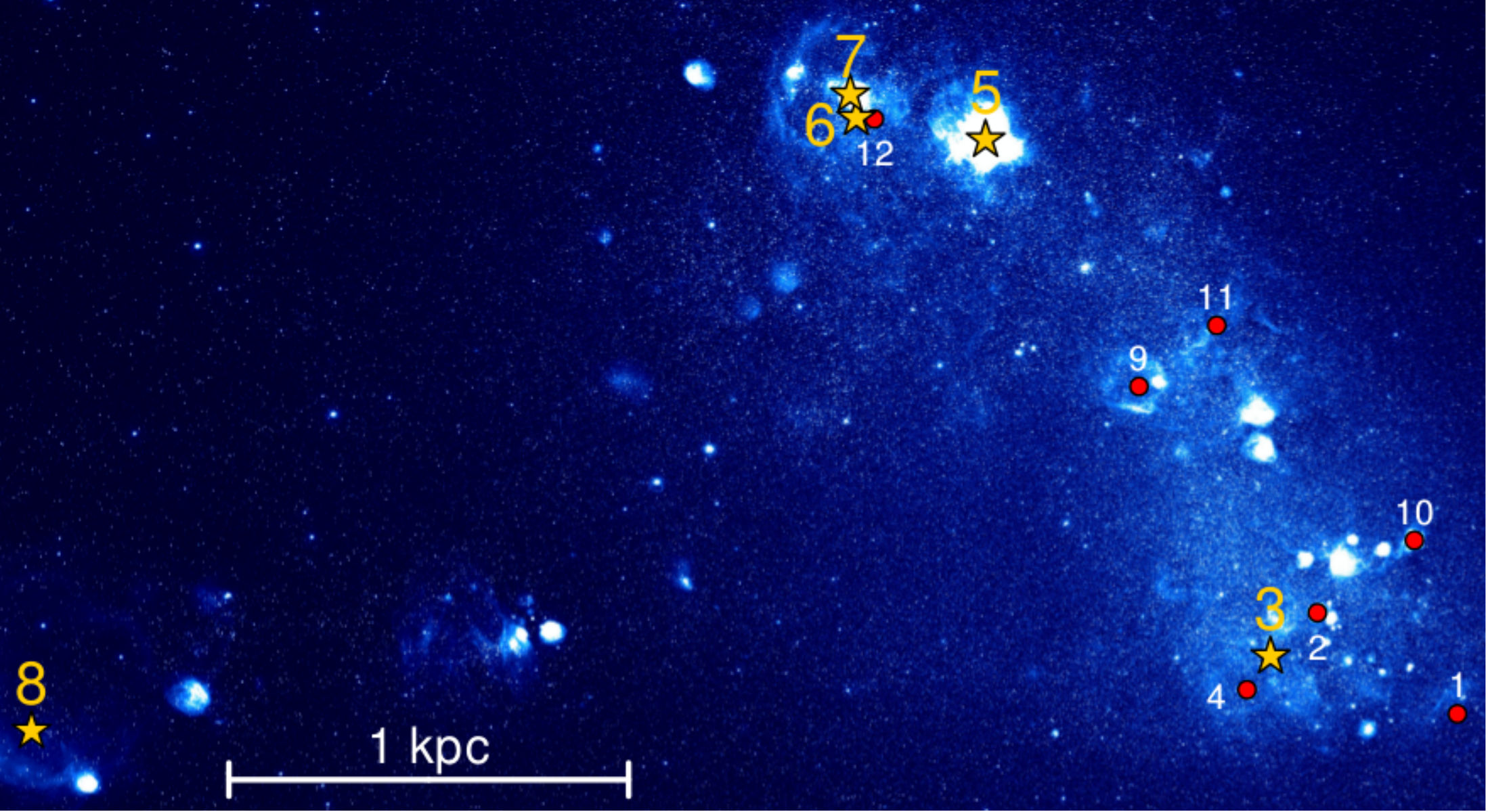}
  \caption{Narrow band O\,[{\sc iii}] nebular emission image of the SMC \citep{Smith2005} with all known WR stars marked. Yellow stars correspond to confirmed binary systems.}
\label{fig:SMCpic}
\end{figure*}

Stars with initial masses $M \gtrsim 20\,M_\odot$ may
reach the Wolf-Rayet (WR) phase, which is characterized by strong stellar winds and 
hydrogen depletion, at a late stage of their evolution as   
they approach the Eddington limit \citep{Conti1976}. There 
are two prevalent channels for a star to do so.  On the one hand,
the powerful radiation-driven winds of massive stars
can peel off their hydrogen-rich
layers, which leads to the typical emission line WR spectrum
\citep{CAK1975, Cassinelli1979a}. On the other hand, 
mass donors in binary systems may shed copious amounts of mass 
during Roche lobe overflow (RLOF),  approaching the 
Eddington limit owing to severe mass-loss 
\citep[e.g.][]{Paczynski1973}. Several studies
 \cite[e.g.][]{Maiz2010, Sota2011, Sana2012, Chini2012, Aldoretta2015} give direct evidence that 
at least half of all massive stars are binaries. 
Among the WR stars, 
about $40\,\%$ are confirmed binaries \citep{Vanderhucht2001}.
It is inevitable that some of these systems would contain interacting companions. 
\cite{Sana2013} estimate that roughly half of 
the O-type stars will interact with a companion via mass transfer during their lifetime, and 
recent studies invoke binary interaction to explain a multitude
of phenomena  \citep[e.g.][]{Vanbeveren2007, Richardson2011, Langer2012, DeMink2013}.
Yet the impact of binarity on the WR population  remains debated \citep[e.g.][]{Vanbeveren1998, Crowther2007}.

Binaries are not only important from an evolutionary standpoint; they further 
enable one to deduce stellar parameters to an accuracy  unattainable for single stars. 
For instance, if the orbital inclination $i$ and both radial velocity (RV) curves 
can be obtained, the companions' masses can be 
accurately calculated from Newtonian dynamics. This method is indispensable in the case of WR stars,
whose masses are otherwise difficult to determine.
Knowledge of these masses provides a critical test, not only of stellar evolution models, but also 
of mass-luminosity relations (MLRs) for WR stars \citep{Langer1989, Graefener2011}. 
Studies of wind-wind collisions (WWC) in massive binaries
have proven fruitful for obtaining orbital inclinations
\citep{Luehrs1997, Moffat1998, Reimer2009}.
These types of wind collisions were also suggested to be prodigious X-ray sources 
\citep{Cherepashchuk1976,Prilutskii1976}, which was fully confirmed by subsequent observations and modeling efforts \citep[e.g.][]{Stevens1992, Zhekov2012,rauw2015}, 
and yielding 
important physical constraints on WR binaries.

Hence, there are various reasons why 
a spectroscopic and photometric analysis of WR binaries is essential for the study of massive stars. Despite this, 
binaries have often been left out in previous spectroscopic studies of WR 
stars \cite[e.g.][]{Hamann2006, Sander2012, Hainich2014}
because of the complexity involved in their analysis. This paper begins to bridge the gap by presenting a systematic 
analysis of WR binaries.

An interesting test case for the impact of binarity on WR stars is offered by investigating low metallicity environments.
Since the mass-loss rate $\dot{M}$ scales with surface opacity that originates in metals, it is expected
to decrease with decreasing surface metallicity $Z$ \citep{Kudritzki1987, Puls2000, Vink2000}. For WR stars, recent empirical 
studies suggest $\dot{M} \propto Z$ \citep[e.g.][]{Nugis2007, Hainich2015}. 
The smaller $Z$ is, the harder it is for a single star 
to develop the stellar wind necessary to become a WR star. Standard stellar evolution models 
predict that, at solar metallicity, 
initial masses of $M_\text{i} \gtrsim 20\,M_\odot$ are 
sufficient for single stars to reach the WR phase, while at a metallicity of about $0.1\,Z_\odot$, 
masses of at least $\sim 45\,M_\odot$ are required \citep{Meynet2005}.
The frequency of single WR stars is thus 
expected to decrease with $Z$.
In contrast, the frequency of WR stars formed via RLOF 
is not a priori expected to depend on $Z$. 

Motivated by such predictions, \citet{Foellmi2003LMC}, \citet*[][FMG hereafter]{Foellmi2003SMC},
\cite{Schnurr2008}, and \citet{Bartzakos2001} conducted 
a large spectroscopic survey in the Small and Large Magellanic Cloud (SMC and LMC, respectively)
with the goal of measuring the binary fraction in their respective WR populations. The LMC and SMC are 
both known to have a subsolar metallicity: a factor $\sim 1/3$ and 
$\sim 1/7$ solar, respectively \citep{Dufour1982, Larsen2000}. 
Following the reasoning of the previous paragraph, 
it is expected that the fraction of WR stars formed 
via RLOF
will be relatively large in the LMC, and even larger in the SMC. 
\citet{Bartzakos2001} made use of stellar evolution statistics published by \cite{Maeder1994}
to predict that
virtually \emph{all} WR stars in the SMC are expected to have been formed via RLOF. Similar predictions remain even 
with the most recent generation of stellar evolution codes \citep[e.g.][]{Georgy2015}. It was therefore surprising that 
FMG measured a WR binary fraction in the SMC of $\sim 40\,\%$, consistent with the Galactic fraction, 
revealing a clear discrepancy between theory and observation which must be explained.

\renewcommand{\arraystretch}{1.3}

\begin{table*}[!htb]
\small
\caption{Overview of the known WR stars in the SMC and their binary status} 
\label{tab:overview}
\begin{center}
\begin{tabular}{lcccccccccc}
\hline 
Object                     & Spectral Type                   &  V\,[mag]        & Binary status                  & $K_1$\,[\kms]&  $K_2$\,[\kms]& $P$ [d]&  $i$ [$^\circ$]\tablefootmark{a} & $e$  \\
\hline                                                                                  
      \object{SMC\,AB\,1}                              & WN3ha                     &   15.1         & -                        &    -         &       -      & -       &      -                  & -   \rule{0mm}{4mm}    \\           
      \object{SMC\,AB\,2}                              & WN5ha                     &   14.2         & -                        &    -         &       -      & -       &      -                  & -       \\           
      {\large \object{SMC\,AB\,3}}\tablefootmark{b}     & WN3h + O9                 &    14.5        & SB2                      &   144        & - & 10.1    &      $57_{-29}^{+33}$\,\tablefootmark{c}    & 0.09    \\ 
      \object{SMC\,AB\,4}                              & WN6h                      &   13.3         & -                        &    -         &       -      & -       &     -                   &  -      \\ 
      {\large \object{SMC\,AB\,5}}\tablefootmark{d} & (WN6h + WN6-7) + (O + ?)&   11.1         & SB3\tablefootmark{e}  &   214        &      200     & 19.3                             & $86^{+1}_{-1}$\,\tablefootmark{f}    & 0.27    \\ 
      {\large \object{SMC\,AB\,6}}             & WN4 + O6.5I:              &   12.3         & SB2                      &   290        &      66      & 6.5     & $57^{+33}_{-12}$\,\tablefootmark{g}  & 0.1     \\
      {\large \object{SMC\,AB\,7}}\tablefootmark{h}    & WN4 + O6I(f)              &   12.9         & SB2                      &   196        &      101     & 19.6    & $68^{+22}_{-15}$\,\tablefootmark{g}     & 0.07    \\
      {\large \object{SMC\,AB\,8}}\tablefootmark{i}    & WO4 + O4V                 &   12.8         & SB2                      &   176        &      55      & 16.6                  & $40^{+10}_{-3}$    & 0       \\
      \object{SMC\,AB\,9}                              & WN3ha                     &   15.2         & Uncertain                &   43      &      -       & 34.2                                 & -   & 0.22 \\    
      \object{SMC\,AB\,10}                             & WN3ha                     &   15.8         & -                        &   -          &      -       & -       &      -                  & -       \\  
      \object{SMC\,AB\,11}                             & WN4ha                     &   15.7         & -                        &   -          &      -       & -       &      -                  & -       \\  
      \object{SMC\,AB\,12}\tablefootmark{j}            & WN4(?)                    &   15.5         & -                        &   -          &      -       & -       &      -                  & -       \\  
\hline
\end{tabular}
\tablefoot{
All entries adopted from FMG unless otherwise specified. Stars analyzed here are denoted with large fonts. \\ 
\tablefoottext{a}{Errors on $i$ are subject to realistic constraints on the O-companion mass (see text)}
\tablefoottext{b}{although AB\,3 is an SB2 binary, FMG could not measure $K_2$ because of the secondary's faintness}
\tablefoottext{c}{$i$ corresponds to a mean value (see text) }
\tablefoottext{d}{spectral type of the primary variable (WN3/11), but WN6h (FMG) corresponds to the most recent spectra used here. All other entries adopted from \citet{Koenigsberger2014}, and references therein}
\tablefoottext{e}{i.e.\ three out of the four components are apparent in the available spectra}
\tablefoottext{f}{adopted from \cite{Perrier2009}}
\tablefoottext{g}{$i$ calibrated against secondary's spectral type (see text)} 
\tablefoottext{h}{orbital parameters adopted from \cite{Niemela2002}, whose solution is more accurate than that presented by FMG (FMG)} 
\tablefoottext{i}{\citet{Moffat1990}, \citet{St-louis2005}, and references therein} 
\tablefoottext{j}{\citet{Massey2003}}
    } 
\end{center}
\end{table*}

\citet[][paper I hereafter]{Hainich2015} conducted a spectral analysis of the putatively single WR stars 
in the SMC. In the present paper, we perform a non-local thermodynamic equilibrium (non-LTE) analysis of the WR binary systems. 
 Having derived the full set of parameters for all components of each 
system, we test current MLRs, and 
deduce  evolutionary paths for each system.

The paper is structured as follows: Sects.\,\ref{sec:sample} and \ref{sec:data} give an overview 
of our sample and the observational data used. In Sect.\,\ref{sec:analysis}, we describe the
assumptions and methods involved in the spectral analysis. In Sect.\,\ref{sec:results}, we 
give the full set of stellar parameters derived, while Sect.\,\ref{sec:disc} contains a thorough 
discussion and interpretation of our results.
A summary of our results is found in Sect.\,\ref{sec:summary}. The appendices
include a detailed description 
on the properties and analysis of our objects (A), their 
evolutionary paths (B), and their spectral fits (C).

\section{The sample}
\label{sec:sample}

There are 12 WR stars currently known in the SMC 
\citep{Massey2014}.
Fig.\,\ref{fig:SMCpic} marks the positions of the known SMC WR stars on a narrow band image of the O\,[{\sc iii}] nebular 
emission line. 
Five out of the 12 stars are in confirmed binary or multiple systems based on their RV curves (FMG),
marked  with yellow stars in Fig.\,\ref{fig:SMCpic}.
As in paper I, we follow the name scheme SMC\,AB\,\# (sometimes simply AB\,\#), as introduced 
by \citet{Azzopardi1979}. 
Table\,\ref{tab:overview} gives an overview on the 12 WR stars currently known in the SMC and their 
binary status.  All but one are WN (nitrogen rich) stars; the WO (oxygen
rich) primary in AB\,8 is the exception.
For the binary systems, we give the two velocity amplitudes $K_1$ and $K_2$, periods $P$, orbital inclinations 
$i$, and 
eccentricities $e$, if known. 
For the quadruple system AB\,5, the orbital parameters refer to the short period WR + WR binary 
within the system. 

The adopted orbital inclination angles $i$  
heavily affect the deduced orbital masses $M_\text{orb}$ ($M_\text{orb} \propto \sin^3 i$). 
The inclinations of AB\,5 and 8 could be constrained in other studies thanks to photospheric/wind 
eclipses. 
Since 
$i$ is the only free parameter 
determining the orbital masses in the case of AB\,6 and 7, we use calibrations 
by \cite{Martins2005} to fix their inclinations so that the secondary's
orbital mass  agrees with its spectral type\footnote{While the calibration depends on $Z$, its effect on the spectral type-mass calibration 
is smaller than the typical errors given here}. With both $i$ and $K_2$ unknown in the case of AB\,3, 
we fix $i$ to its mean statistical value 
so that $\sin^3 i = \left < \sin^3 i \right > = 3\,\pi / 16$, and adjust $K_2$ to calibrate the secondary's mass against its
spectral type.

In all systems, we conservatively assume an uncertainty of no more than a factor two in the secondary's orbital mass, 
and assume the realistic constraints 
$15\,M_\odot \le M_\text{O} \le 70\,M_\odot$ and $ M_\text{WR} \ge 5\,M_\odot$ for the O and WR stars in the sample, respectively \citep[e.g.][]{Martins2005, Crowther2007}.
If additional constraints on $i$ are known from other studies,
they are considered as well. 
This restricts $i$ to a corresponding value range.

The binary candidate SMC\,AB\,9  (cf.\ Table\,\ref{tab:overview}) is omitted from this analysis. The star was already
analyzed as a single star in \mbox{paper I} because of the absence of any spectral features which could be 
associated with a companion in its spectrum. 
For the same reason,
we treat SMC\,AB\,5 (HD 5980) as a triple system, and not a quadruple. None of the spectral features 
are clearly associated with the fourth component, 
whose existence is anticipated based on a periodic variability of the absorption lines associated with the third component \citep{Breysacher1982, Koenigsberger2014}.

\section{Observational data}
\label{sec:data}

\subsection{Spectroscopic data}
\label{subsec:specdata}


For three systems (AB\,3, 6, and 7), we use normalized, low-resolution spectra ($\text{FWHM}\approx$ 2.8 -- 6.7\,\AA) obtained by FMG in the spectral 
range of 3900--6800\,$\AA$  between the years 1998 and 2002.
Detailed information on the instrumentation used and the data reduction can be found in FMG.
To obtain a relatively high Signal-to-Noise ratio (S/N) of about $100-150$, 
spectra taken at different binary phases were co-added in the frame of the primary.
Although the reduced spectra used by FMG for RV studies are available for download, the online data suffer from 
obvious wavelength calibration problems for reasons that we could not trace. The original data could not be retrieved. 
Due to this and the poor S/N of the original spectra, 
we make use only of the co-added spectra in this study. Co-adding the spectra in the frame of the WR star causes the companion's spectral
features (an O star in these systems) to smear. Its lines would thus appear broader and shallower than they should, although their 
equivalent width remains conserved. To account for this effect, we convolve the companion's model with a 
box function of the width $K_1 + K_2$ when comparing to these spectra. Given the low resolution of the spectra, this effect is 
of secondary importance. If possible, auxiliary spectra were used to derive parameters which are sensitive to the line profile.

For all of our targets, we downloaded flux calibrated spectra taken with the International Ultraviolet Explorer (IUE) 
covering the spectral range $1200 - 2000\,\AA$ from the MAST archive. 
When available, high resolution spectra are preferred, binned at intervals of $0.05\,\AA$ to achieve an S/N$\approx 15$. Otherwise, 
low resolution spectra are used ($\text{FWHM} \approx 6\,\AA$, $\text{S/N} \approx 20$). Low resolution, flux calibrated IUE spectra in the range $2000 - 3000\,\AA$ are not used 
for detailed spectroscopy because of their low S/N ($\approx 5-10$), but rather to cover the spectral energy distribution (SED) of the targets.
Optical low resolution spectra taken by \citet{Torres1988} are also used for the SEDs of our targets.
Flux calibrated, high resolution Far Ultraviolet Spectroscopic Explorer (FUSE) spectra covering the spectral range $960-1190\,\AA$
are also retrieved from the MAST archive and binned at $0.05\,\AA$ to achieve an $\text{S/N}\approx 30$, except for AB\,3, for which no 
usable FUSE spectra could be obtained.
The IUE and FUSE spectra are normalized with the reddened model continuum.

For AB\,5, we use auxiliary flux calibrated, high resolution ($\text{FWHM} \approx 0.02\,\AA$) spectra taken  in 2009 during primary eclipse ($\phi = 0$) with the STIS 
instrument mounted on the Hubble Space Telescope (HST) covering the spectral range $1150-1700\,\AA$, with $\text{S/N}\approx 30$
\citep{Koenigsberger2010}. The spectra 
are also normalized using the model continuum.  Unfortunately, no out-of-eclipse UV spectra taken after the year 1999 are available in the archives.
For this system, we also retrieved several high resolution spectra taken in 2005 with the Fiber-Fed Extended Range Optical Spectrograph (FEROS) mounted on the 
MPI 2.2\,m telescope at La Silla, covering the spectral range $3750 - 9200\,\AA$ at a resolution of $\text{FWHM} \approx 0.1\,\AA$ and $\text{S/N}\approx 40$. 
The data were reduced and used by \cite{Foellmi2008}, where more information can be found. 
Additionally, we use a spectrum created by co-adding several spectra at phase $\phi = 0$ to 
achieve a $\text{S/N}\approx 200$, made available by \cite{Foellmi2008}. 

For AB\,8, we make use of a reduced, flux calibrated spectrum with a resolution of $\text{FWHM} \approx 0.2\,\AA$ and $\text{S/N}\approx 100$
taken with the X-shooter spectrograph mounted on ESO's Very 
Large Telescope (VLT), covering the range $3000 - 25000\,\AA$ \citep{Vernet2011}. The reduced spectrum was kindly
supplied to us by F.\ Tramper (see \cite{Tramper2013} for details). 

The dates and ID numbers of all spectra used here are given in the figures showing them. 
Phases are calculated with ephemeris given by FMG, except for AB\,8 and 5, where the 
ephemeris given by \cite{St-louis2005} and \cite{Koenigsberger2014} 
are used, respectively. The spectral resolution
is accounted for by convolving the models with corresponding Gaussians to mimic the instrumental profile.

\subsection{Photometric data}
\label{subsec:photdata}

Aside from flux-calibrated data, we use for all stars analyzed here
$JHK$ and IRAC photometry from \citet{Bonanos2010}. If available, we also use their $UBVRI$ magnitudes. 
For AB\,5, we use the $U$ and $BVR$ magnitudes from  \cite{Torres1988} and \cite{Zacharias2005}, respectively; 
For AB\,6, we use UBV 
magnitudes from \cite{Mermilliod1995}; for AB\,8, we use the 
UBV magnitudes from \cite{Massey2002} and $I$ magnitude given by the \cite{DENIS2015}. 
For all stars, we use WISE magnitudes 
from \cite{Cutri2013}.

\section{Non-LTE spectral modeling of WR binaries}
\label{sec:analysis}

\subsection{The PoWR code}
\label{subsec:code}

PoWR is a non-LTE model atmosphere
code especially suitable for hot stars with expanding atmospheres\footnote{PoWR models of Wolf-Rayet stars can be downloaded at
http://www.astro.physik.uni-potsdam.de/PoWR.html}. The code iteratively solves the co-moving frame
radiative transfer and the statistical balance equations in spherical symmetry under the constraint
of energy conservation. 
A more detailed description of the assumptions and methods used
in the code is given by \citet{Graefener2002} and \citet{Hamann2004}. By
comparing synthetic spectra generated by the code to observations, a multitude of stellar
parameters can be derived.

The inner boundary of the model, 
referred to as the stellar radius $R_*$,
is defined at the Rosseland optical depth $\tau_\mathrm{Ross}$=20, where LTE can be safely assumed. 
In the subsonic region, the velocity
field is defined so that a hydrostatic density stratification is
approached \citep{Sander2015}.
In the supersonic region, the
pre-specified wind velocity field $v(r)$ generally takes the form of the $\beta$-law \citep{CAK1975}

\begin{equation}
 v(r) = v_\infty \left(1 - \frac{r_0}{r}\right)^{\beta}.
\label{eq:2betalaw}
\end{equation}
Here, $v_\infty$ is the terminal velocity, 
and $r_0 \approx R_*$ is a constant determined so as
to achieve a smooth transition between 
the subsonic and supersonic regions.

Beside the velocity law and chemical composition, 
four fundamental input parameters are needed to define a model atmosphere: 
the effective temperature $T_*$,
the surface gravity $g_*$, the mass-loss rate $\dot{M}$, and the stellar luminosity $L$.
The effective temperature relates to $R_*$ and $L$ via 
the Stefan-Boltzmann law: $L = 4\,\pi\,\sigma\,R_*^2\,T_*^4$.  
The gravity $g_*$ relates to the radius $R_*$ and mass $M_*$  via the usual definition: $g_* = g(R_*) = G\,M_* R_*^{-2}$. 
For the vast majority of WR models, the value of $g_*$ bears no significant effects on the synthetic spectrum, 
which originates primarily in the wind.  
The outer boundary is taken to be 
$R_\text{max} = 100\,R_*$  for O models and $1000\,R_*$ for WR models, 
which were tested to be sufficient.  

During the iterative solution, the line opacity and emissivity profiles at each radial layer are 
Gaussians with a constant Doppler width $v_\text{Dop}$. This parameter
is set to $30$ and $100\,$\kms\, for O and WR models, respectively.
In the formal integration, the
Doppler velocity is decomposed to depth-dependent thermal motion and microturbulence $\xi(r)$. 
We assume $\xi(r)$ grows with the wind velocity up to
\mbox{$\xi(R_\text{max}) = 0.1\,v_\infty$}, and set $\xi(R_*) = 20$\,\kms\, for O models 
and 100\,\kms\, for WR models, respectively \citep[e.g.][]{Hamann2006, Bouret2012, Shenar2015}.
We assume a macroturbulent velocity 
of $30\,$\kms\, for all O components \citep[e.g.][]{Markova2008, Bouret2012}, accounted for by 
convolving the profiles with radial-tangential profiles \citep[e.g.][]{Gray1975, SimonDiaz2007}.

It has become a consensus that winds of massive stars are not smooth, but 
rather clumped
\citep{Moffat1988, Lepine1999, Markova2005, Oskinova2007, Prinja2010, Surlan2013}.
An approximate treatment of optically thin clumps using the so-called microclumping approach was introduced by 
\cite{Hillier1984} and systematically implemented by \cite{Hamann1998}, 
where the population numbers are calculated in clumps which are a factor of $D$ denser 
than the equivalent smooth wind ($D = 1 / f$, where $f$ is the filling factor). 

Because optical WR spectra  are dominated by recombination lines, 
it is customary 
to parametrize their models using the so-called transformed radius \citep{Schmutz1989},

\begin{equation}
 R_\text{t} = R_* \left[ \frac{v_\infty}{2500\,{\rm km}\,{\rm s}^{-1}\,}  \middle/  
 \frac{\dot{M} \sqrt{D}}{10^{-4}\,M_\odot\,{\rm yr}^{-1}}  \right]^{2/3},
\label{eq:Rt}
\end{equation}
defined such that equivalent widths of recombination lines of models  
with given $R_\text{t}$ and $T_*$ are 
approximately preserved, independently 
of $L$, $\dot{M}$, $D$, and $v_\infty$. While $R_\text{t}$ has the dimensions of length, it should be 
thought of as an integrated volume emission measure per stellar surface area.

\subsection{Assumptions}
\label{subsec:assumptions}

PoWR models are limited to spherical symmetry, which obviously breaks down in the case 
of close binary systems. Firstly, the stars are deformed into a tear-like 
shape due to tidal forces. Secondly,  non-spherical  manifestations resulting from binary 
interaction, such as WWC cones or
asymmetrical accretion flows, may occur in binary systems. While such phenomena may be significant or even 
dominant in the case of specific lines \cite[e.g.][]{Bartzakos2001b}, 
they typically amount to flux variations of the order a few percent \cite[e.g.][]{Hill2000, Palate2013}, 
with the possible exception of AB\,5 (see Appendix\,\ref{sec:comments}). 

The detailed form of the velocity field in the wind domain can affect spectral 
features originating in the wind. WR models are therefore more sensitive to the velocity 
law than O-star models. The finite disk correction 
predicts a $\beta$-law (Eq.\,\ref{eq:2betalaw}) in the case of OB-type stars with $\beta = 0.8$ \citep[e.g.][]{Kudritzki1989},  which we 
adopt for the O-star models. This 
value is consistent with analyses of clump propagation in 
O-type stars \citep[e.g.][]{Eversberg1998}. 
As for WR stars, there are several empirical studies which suggest 
that the $\beta$ exponents in the outer winds of WR stars with strong winds 
are in excess of four \citep{Lepine1999, Dessart2005}. 
On the other hand, 
$\beta$ values of the order of unity are found for hydrogen 
rich WR stars \citep{Chene2008}, which describes most of our sample.
For the WR components, we thus always assume the usual $\beta$-law with $\beta = 1$. 

The data do not always enable us to derive the chemical abundances for each element, in which 
case we assume the following.
For the O companions,
we adopt H, C, N, O, Mg, Si, and Fe mass fractions as derived for B stars in the SMC by \cite{Korn2000}, \cite{Trundle2007}, and \cite{Hunter2007}: 
$X_\text{H} = 0.73$, $X_\text{C} = 2.1\cdot10^{-4}$, $X_\text{N} = 3.26\cdot10^{-5}$, 
$X_\text{O} = 1.13\cdot10^{-3}$, $X_\text{Mg} = 9.9\cdot10^{-5}$, $X_\text{Si} = 1.3\cdot10^{-4}$, 
and $X_\text{Fe}  = 3\cdot10^{-4}$.
We scale the mass fractions of the remaining metals to 1/7 solar, in accordance with 
the ratio of the solar metallicity \citep{Asplund2009} to the average SMC metallicity \citep{Trundle2007}: 
$X_\text{Ne} = 1.8\cdot10^{-4}$, $X_\text{Al} = 7.6\cdot 10^{-6}$, $X_\text{P} = 8.3\cdot 10^{-7}$, 
$X_\text{S} = 4.4\cdot10^{-5}$. The same mass fractions are adopted for WR models, except for $X_\text{H}$, which is derived in each case,  
and the CNO mass fractions, which are adopted as in paper I: 
$X_\text{C} = 2.5\cdot10^{-5}$, $X_\text{N} = 1.5\cdot10^{-3}$, 
$X_\text{O} = 1.0\cdot10^{-6}$.

A longstanding problem is assessing the density contrast in clumps and their
stratification in the atmosphere.
Here, $D(r)$ is assumed to be depth-dependent, 
increasing from $1$ (smooth wind) at the base of the wind to a maximum value $D$ \citep[e.g.][]{Owocki1988}.
Quite generally, allowing $D(r)$ to be depth-dependent yields more symmetric, 
less ``self-absorbed'' line profiles, as illustrated in Fig.\,\ref{fig:DDclump} in the appendix. It is found that 
a depth dependent clumping factor which initiates from $D(R_*) = 1$ at the base of the wind 
and grows proportionally to the wind velocity to the value  $D(r) = D$ at $v(r) \ge 0.5\,v_\infty$ provides a good agreement 
with the observations, which tend to exhibit symmetric profiles.
The maximal value $D$ can 
be roughly constrained for each WR star (see Sect.\,\ref{subsec:method}), and is treated as a free parameter.  
We note that
other studies suggest clumping may already initiate at the photosphere
\citep[e.g.][]{Cantiello2009, Torrejon2015}.
Clumping factors 
for the O companions, which cannot be deduced from the available spectra, are
fixed to $D = 10$, supported by hydrodynamical simulations \citep{Feldmeier1997}. 
When the companions' mass-loss rates cannot be constrained,
we adopt them from hydrodynamical predictions by \citet{Vink2000}.

We adopt a distance of $d = 62$\,kpc to the SMC \citep{Keller2006}. 
The reddening towards our objects is modeled using a combination of the reddening laws derived by \cite{Seaton1979} 
for the Galaxy,
and by \cite{Gordon2003} for the SMC. As in paper I, we assume an extinction of $E_{B-V} = 0.05$ for the Galactic component (Sect.\ 4.3 in paper I) and fit 
for the total 
extinction, adopting $R_\text{V} = 3.1$.

\subsection{The analysis method}
\label{subsec:method}

The non-LTE analysis of spectra, even in the case of single stars, is an iterative and
computationally expensive process.
Generally, $g_*$ 
is inferred from the wings of photospheric 
H and He\,{\sc ii} absorption lines. The effective temperature $T_*$
is inferred from the line ratios  
of ions belonging to the same element, mostly He lines for the O stars, and mostly metal lines for 
WR stars. Wind parameters such as $\dot{M}$ (or $R_\text{t}$) and $v_\infty$ are derived
from recombination and P Cygni lines. If possible, the maximum density contrast $D$ 
is derived from electron-scattering 
wings. The luminosity $L$ and total extinction $E_{B-V}$ are determined 
by fitting the combined spectral energy distribution (SED) of the models to the photometric 
measurements.
The abundances are determined from the overall
strengths of lines belonging to the respective elements. 
Finally,  the projected rotation velocity $v \sin i$
is constrained from profile shapes. For the O companions, this is done by 
convolving the models with appropriate rotation profiles. For the WR stars, if the 
resolution and S/N enable such an analysis,
we derive upper limits for rotation by applying a 3D integration scheme, assuming co-rotation up to $\tau_\text{Ross} = 2/3$
\citep{Shenar2014}.

\begin{figure}[!htb]
\centering
  \includegraphics[width=\columnwidth]{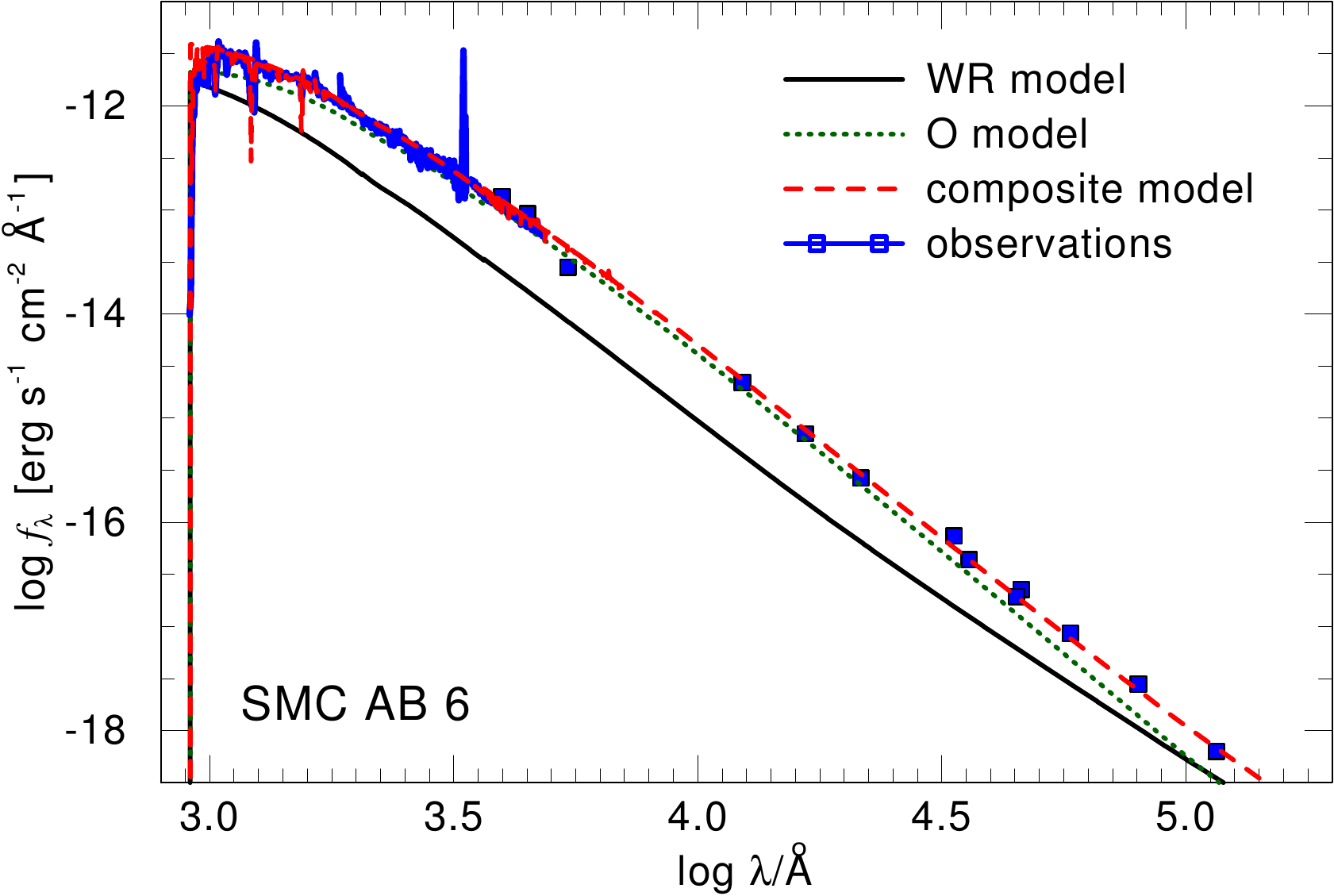}
  \caption{Observed SED of AB\,6 (blue lines and squares) compared to the synthetic SED (red dashed line), 
    which is the sum of the WR (black solid line) and O (green dotted line) 
    models.} 
\label{fig:WR6SED}
\end{figure}

\begin{figure*}[!htb]
\centering
  \includegraphics[width=\textwidth]{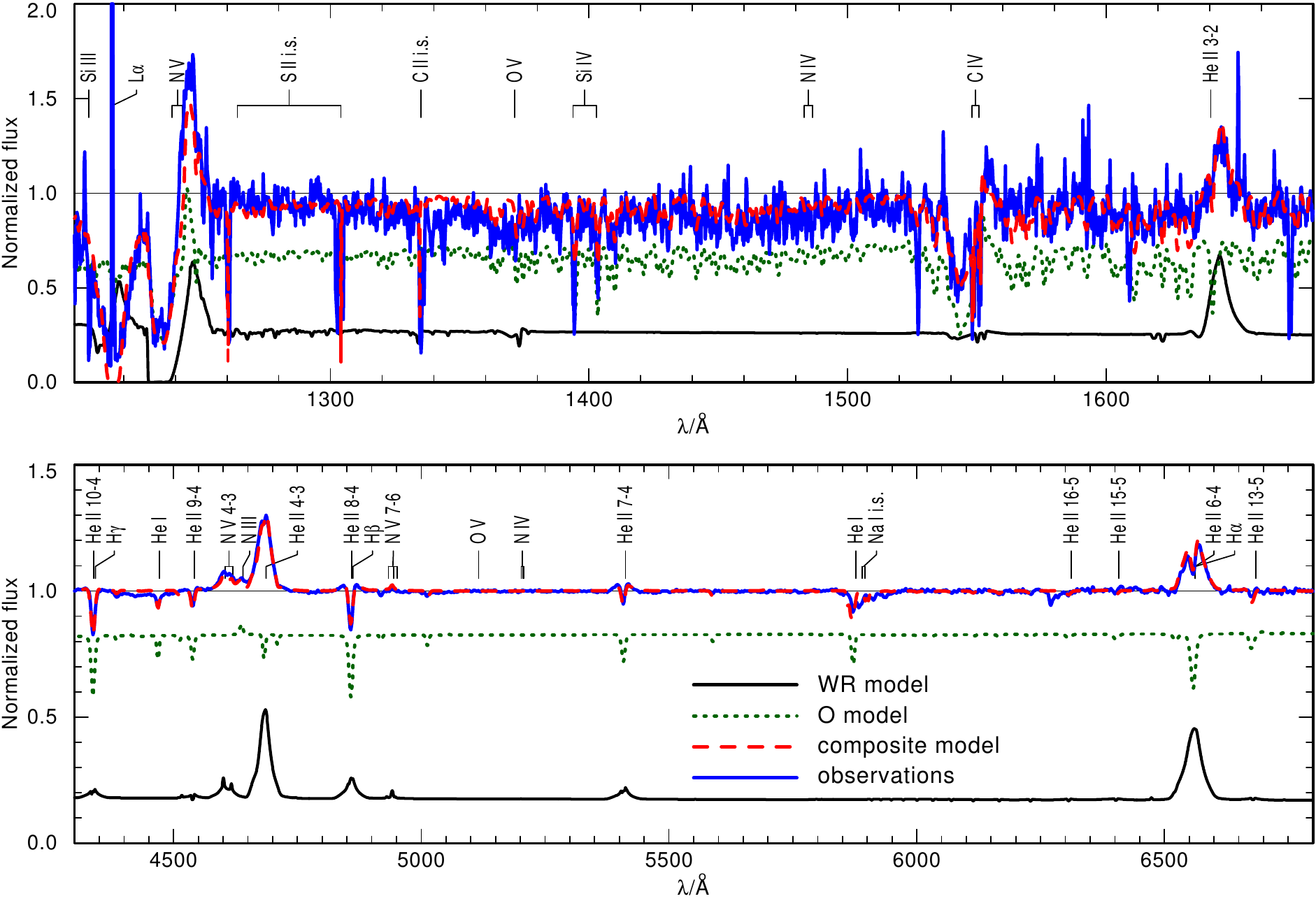}
  \caption{Comparison between IUE (ID:sp41784, $\phi = 0.47$) and optical  (co-added, FMG) rectified spectra of SMC\,AB\,6 (blue ragged line) and the composite synthetic spectrum (red dashed lines). The composite model 
    is the sum of the WR (black solid line) and O (green dotted line) models. The relative offsets of the model continua correspond to the light ratio between the two stars. Note 
    that the light ratios are different in the optical and UV due to the different temperatures of both components.}
\label{fig:WR6spec}
\end{figure*}

To analyze a multiple system, models for each of its components are required. 
Ideally, one would disentangle the composite spectrum to its constituent spectra 
by observing the system at different phases \cite[e.g.][]{Bagnuolo1991, Hadrava1995, Marchenko1998}.
Unfortunately, our data do not enable this. Moreover, spectral disentangling does not yield direct information
regarding the light ratios unless the stellar system is eclipsing.
Since we work with 
composite spectra, our task is therefore to combine models
in such a way that the composite spectrum and SED are reproduced. An example is shown in Figs.\,\ref{fig:WR6spec} and \ref{fig:WR6SED}, where 
a comparison between the  SED and observed rectified spectra of the binary system SMC\,AB\,6 
and our best fitting models is shown, respectively.

As opposed to single stars, the luminosities of the components 
influence their relative contribution to the flux and thus  the 
synthetic normalized spectrum.
The light ratios of the different components therefore become entangled with the 
fundamental stellar parameters, and it is not trivial to overcome the resulting parameter degeneracy.
The analysis of composite spectra thus consists of the following steps: 

\begin{itemize}
\item {\it Step 1:} Based on line ratios and previous studies (e.g. spectral types), preliminary models 
      for the O and WR companions are established. 
        If necessary, the spectra are shifted to account for systemic/orbital motion.
\item {\it Step 2:} The light ratios are derived (or constrained) by identifying absorption features which can be clearly associated with the O companion and 
      which are preferably not sensitive to variations of its physical parameters.
  While identifying the WR lines is usually 
easier, their strengths strongly depend on the mass-loss rate and thus do not enable one to 
determine the light ratios independently.
\item {\it Step 3:} The luminosity of one of the companions and the reddening of the system are 
        adjusted to fit the available photometry. Since the light ratio is known/constrained, 
        the luminosity of the companion follows.
\item {\it Step 4:} $\dot{M}$ (or $R_\text{t}$), $D$, and $v_\infty$ are adjusted for the WR model based on the strengths of its lines.
\item {\it Step 5:} If needed, the parameters of the WR and O models are further refined. If any wind lines 
        can be associated with the O companion, its wind parameters are adjusted.
\item {\it Step 6:} With the refined models, steps 2 - 5 are repeated until 
no significant improvement to the fit of prominent lines (at a few percent level) can be achieved.
\end{itemize}

\begin{figure}[!htb]
\centering
  \includegraphics[width=\columnwidth]{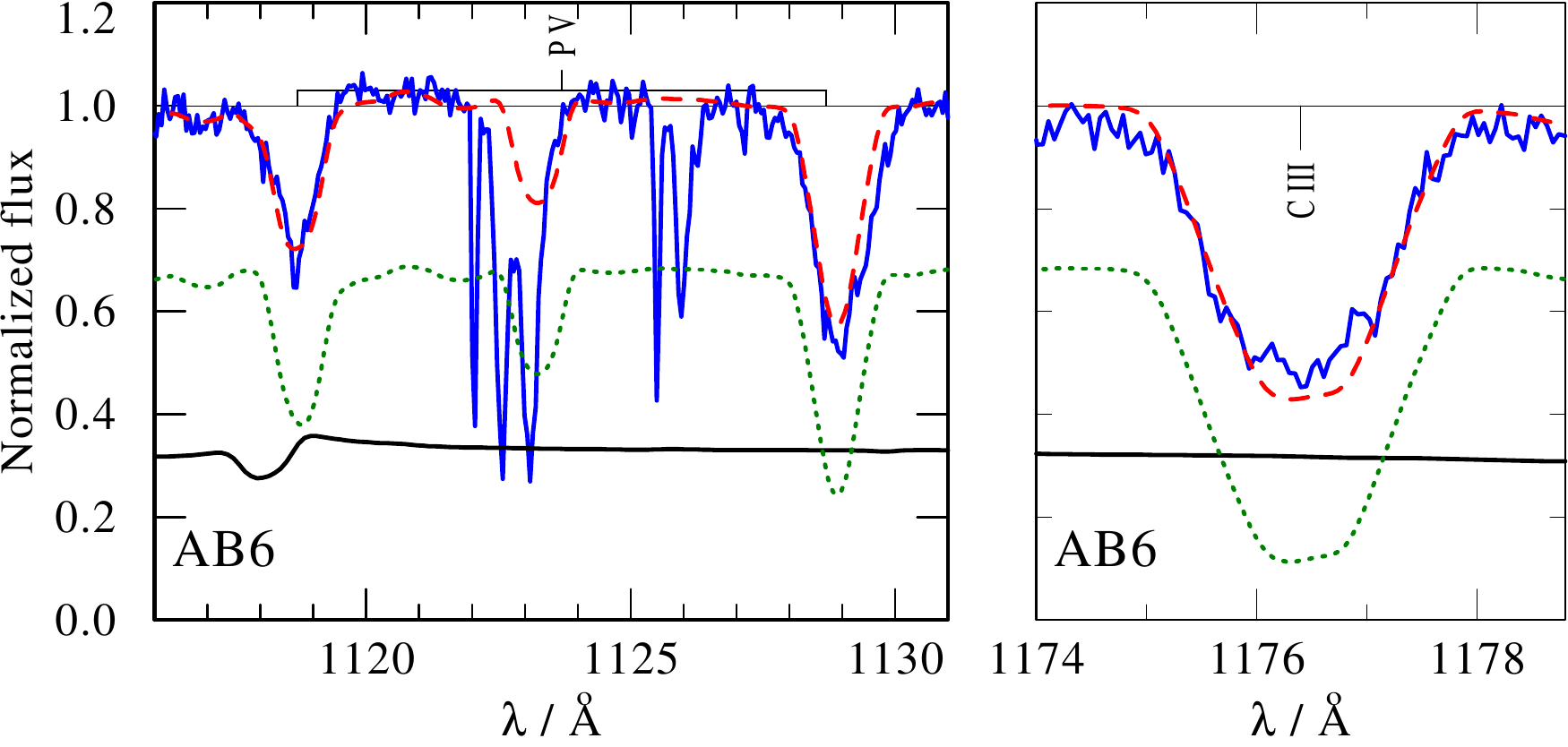}
  \caption{Comparison between the rectified FUSE observation of AB\,6 (ID:X0150102000, $\phi = 0.44$, blue ragged line) and the composite synthetic spectrum (red dashed line) for 
   the  P\,{\sc v} $\lambda \lambda\,1118, 1128$ resonance doublet and the C\,{\sc iii} multiplet at $\sim \lambda 1176$. 
   The WR and O models are depcited by a black solid line and green dotted line, respectively. The narrow absorption features originate 
   in the interstellar medium (ISM).}
\label{fig:CIIIhelp}
\end{figure}

The set of spectral lines most diagnostic for the analysis generally depends on the system.
In Fig.\,\ref{fig:CIIIhelp}, we show an example for two photospheric features which originate in the secondary beyond doubt, and which 
greatly help to deduce the light ratio 
in the case of SMC\,AB\,6: 
the P\,{\sc v} $\lambda \lambda\,1118, 1128$ resonance doublet (left panel) and the strong C\,{\sc iii} 
multiplet at $\approx \lambda 1176$ (right panel). 
Optical He lines as well as the spectral type imply $30 < T_* < 40\,$kK for the secondary.
A careful comparison of O star models in this temperature range 
reveals that these lines are insensitive to temperature and gravity variations.
We thus conclude that the relative strength of such lines in the normalized spectrum is affected primarily by the light ratio.
The features imply a similar light ratio of $F(\text{O}) / F(\text{WR}) \sim 2$ in the FUSE domain, 
and agree with the other features in the available spectra,  e.g. sulfur lines. While this method 
is sensitive to the adopted abundances, 
$X_\text{P}$ and $X_\text{S}$ should remain fairly constant \citep[e.g.][]{Bouret2012} throughout the stellar evolution. 
A multitude of lines is used for each system
to reduce the probability for a systematic deviation.

\begin{figure}[!htb]
\centering
  \includegraphics[width=\columnwidth]{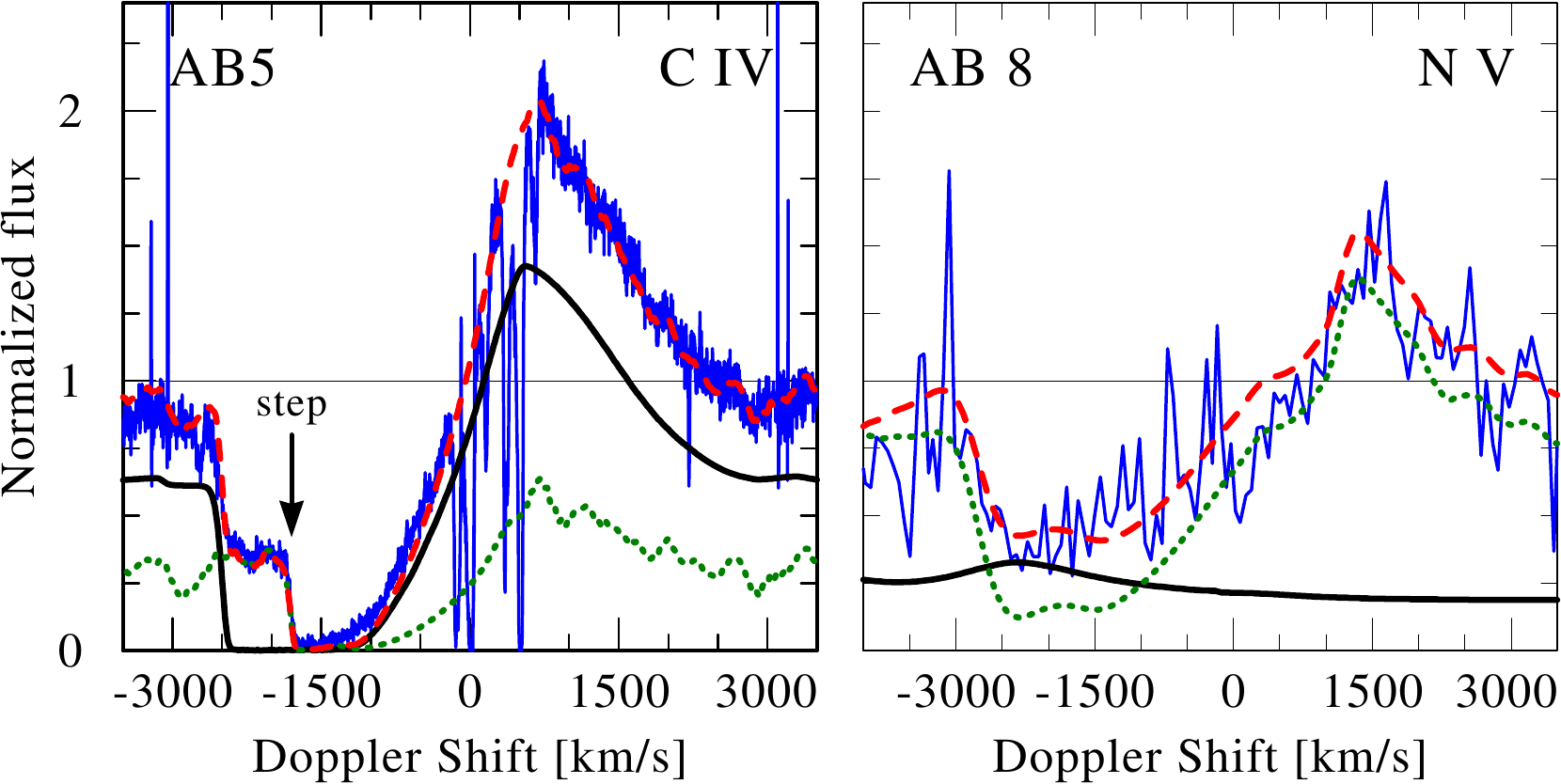}
  \caption{Same as Fig.\,\ref{fig:CIIIhelp}, but showing the observed C\,{\sc iv} $\lambda \lambda\,1548, 1551$ resonance doublet
   in AB\,5 (HST, ID:ob2na1020, $\phi \cong 0$, left panel) 
    and the observed N\,{\sc v} resonance doublet  $\lambda \lambda\,1239, 1243$ in AB\,8 (IUE, ID:sp07623, $\phi = 0.79$, right panel)
    in velocity space relative to the respective blue components of each doublet.
    Colors are as in Fig.\,\ref{fig:CIIIhelp}.}
\label{fig:CIVhelp}
\end{figure}

Another robust way for determining the light ratios is offered by high resolution P Cygni line profiles. An example 
is shown in the left and right panels of Fig.\,\ref{fig:CIVhelp}. 
The left panel shows a high resolution HST spectrum of the C\,{\sc iv} resonance 
doublet of the quadruple system AB\,5, taken at $\phi \cong 0$ during an eclipse of the secondary (B) by the primary (A). The spectrum 
clearly shows a P\,Cygni absorption consisting of two contributions originating in the primary and tertiary (C).
As \citet{Georgiev2011} already demonstrated, the strength of the ``step'' 
observed in the C\,{\sc iv} doublet is influenced by the components' light ratio. The different terminal velocities of stars A and C cause the more extended part of the 
line to appear unsaturated. The right 
panel shows the N\,{\sc v} resonance doublet $\lambda \lambda\,1239, 1243$ for the WO binary AB\,8, which clearly originates 
in the O companion and which is typically saturated in observations of single O stars with strong winds \cite[e.g.][]{Walborn2008, Bouret2012}, 
but is not saturated here because of light dilution by the WO component. 
Such features give sharp constraints on the light ratios.

WR stars are almost always devoid of pure photospheric features and so their surface gravities cannot be 
determined via spectral analysis. Their gravities are fixed to $G\,M_\text{orb}\,R_*^{-2}$ throughout the analysis, after 
$R_*$ has been determined, but we note that the appearance of the WR spectra calculated here are virtually independent of $\log g$.
Determining the gravity of the secondaries proved to be a hard task, leading to large errors in $\log g$.
As described in Sect.\,\ref{sec:data}, 
the co-added optical spectra of AB\,3, 6, and 7  suffer from low resolution and a smearing of the companion's features. However, while the profiles of the 
Balmer and He\,{\sc ii} 
lines cannot be studied in detail because of the quality of the spectra, their equivalent widths grow with increasing $\log g_*$,
which enabled its rough estimation.

In Appendix\,\ref{sec:comments}, we give an overview on each analyzed system, supply a thorough documentation of the analysis, and highlight 
spectral features of notable interest.

\renewcommand{\arraystretch}{1.2}

\begin{table*}[!htb]
\normalsize
\footnotesize
\caption{Inferred stellar parameters for the SMC WR binaries} 
\label{tab:stellarpar}
\begin{center}
\tabcolsep=0.1cm
\begin{tabular}{l | c c | c c c |cc |cc |cc}
\hline
                                              &                 \multicolumn{2}{c|}{AB3}         &                           \multicolumn{3}{c|}{AB5}                             &               \multicolumn{2}{c|}{AB6}          &              \multicolumn{2}{c|}{AB7}         &               \multicolumn{2}{c}{AB8} \\ 
Component & A & B & A & B & C & A & B & A & B & A & B \\
\hline 
Spectral type\tablefootmark{a}                & WN3h                      & O9                   & WN6h                          & WN6-7                    &           O                   & WN4                     & O6.5 I                & WN4                        & O6 I(f)                 & WO4                       & O4 V \rule{0mm}{3mm} \\
$T_{\ast}$ [kK]                               & $78^{+5}_{-5}$            & $30^{+5}_{-5}$       &  $45^{+5}_{-5}$               & $45^{+10}_{-7}$          & $34^{+3}_{-3}$              & $80^{+15}_{-10}$        & $37^{+3}_{-3}$         & $105^{+20}_{-10}$          & $36^{+3}_{-3}$          & $141^{+60}_{-20}$        & $45^{+5}_{-5}$ \\
$T_{2/3}$ [kK]                                & $77^{+5}_{-5}$             & $30^{+3}_{-5}$      &  $43^{+5}_{-5}$             & $43^{+10}_{-7}$        & $33^{+3}_{-3}$            & $78^{+15}_{-10}$        & $37^{+3}_{-3}$         & $98^{+20}_{-10}$           & $35^{+3}_{-3}$        & $115^{+10}_{-10}$        & $44^{+5}_{-5}$ \\
$\log g_*$ [cm\,s$^{-2}$]\tablefootmark{b}    & $4.3$                     & $3.9^{+0.3}_{-0.3}$  &  $3.5$                        & $3.5$                   & $3.2^{+0.2}_{-0.2}$          & $3.7$                   & $3.5^{+0.2}_{-0.2}$    & $4.7$                     & $3.6^{+0.2}_{-0.2}$       & $5.1$                    & $4.0^{+0.3}_{-0.3}$ \\
$\log L$ [$L_\odot$]                          & $5.93^{+0.05}_{-0.05}$     & $4.5^{+0.2}_{-0.2}$   &  $6.35^{+0.10}_{-0.10}$         & $6.25^{+0.15}_{-0.15}$   & $5.85^{+0.10}_{-0.10}$        & $6.28^{+0.10}_{-0.10}$  & $5.90^{+0.10}_{-0.10}$    & $6.10^{+0.10}_{-0.10}$      & $5.50^{+0.10}_{-0.10}$       & $6.15^{+0.10}_{-0.10}$     & $5.85^{+0.10}_{-0.10}$ \\
$\log R_\text{t}$ [$R_\odot$]                 & $1.07^{+0.05}_{-0.05}$     & -                     &  $1.15^{+0.05}_{-0.05}$        & $1.3^{+0.1}_{-0.1}$     & -                           & $1.2^{+0.1}_{-0.1}$   & -                       & $0.75^{+0.10}_{-0.10}$      & -                       & $0.40^{+0.05}_{-0.05}$      & - \\
$v_{\infty}/10^3$ [\kms]                      & $1.5^{+0.1}_{-0.1}$        & $2.0^{+0.5}_{-0.5}$     &  $2.2^{+0.2}_{-0.2}$          & $2^{+0.5}_{-0.5}$        & $1.65^{+0.1}_{-0.1}$        & $2.2^{+0.2}_{-0.2}$    & $2.0^{+0.2}_{-0.2}$      & $1.7^{+0.2}_{-0.2}$       & $1.5^{+0.3}_{-0.3}$      & $3.7^{+0.3}_{-0.3}$      & $3.2^{+0.3}_{-0.3}$ \\
$R_*$ [$R_\odot$]                             & $5^{+1}_{-1}$          & $7^{+6}_{-3}$   &  $24^{+10}_{-7}$             & $22^{+15}_{-10}$       & $24^{+8}_{-6}$             & $7^{+3}_{-2}$       & $22^{+7}_{-5}$          & $3.4^{+1.2}_{-1.2}$        & $14^{+5}_{-3}$        & $2^{+1}_{-1}$             & $14^{+6}_{-4}$ \\
$R_{2/3}$ [$R_\odot$]                         & $5^{+1}_{-1}$          & $7^{+6}_{-3}$   &  $26^{+10}_{-7}$             & $23^{+15}_{-10}$       & $25^{+8}_{-6}$             & $7^{+3}_{-2}$       & $22^{+7}_{-5}$          & $4.0^{+1.4}_{-1.4}$        & $15^{+5}_{-3}$       & $3^{+1}_{-1}$             & $15^{+6}_{-4}$ \\
$D$                                           & $10^{+10}_{-5}$          & $10$                  &  $40_{-20}$                    & $10$                    & $10$                           & $10^{+10}_{-5}$       & $10$                           & $10^{+10}_{-5}$         & 10                    & $40_{-20}$         & 10 \\
$\log \dot{M}$ \smy\tablefootmark{c}         & $-5.3^{+0.1}_{-0.1}$     & $-8.0$                &  $-4.5^{+0.1}_{-0.1}$         & $-4.5^{+0.3}_{-0.3}$      & $-5.9$                    & $-5.1^{+0.2}_{-0.2}$   & $-5.8$               & $-5.0^{+0.2}_{-0.2}$      & $-7.0^{+0.3}_{-0.5}$          & $-4.8^{+0.1}_{-0.1}$       & $-6.1^{+0.5}_{-0.3}$ \\
$v \sin i$ [\kms]                             & -                         & $200^{+100}_{-100}$          &  $< 300$                      & $< 400$                 & $120^{+30}_{-30}$          & -                       & $150^{+50}_{-50}$           & -                      & $150^{+30}_{-30}$          & -                         & $120^{+20}_{-20}$ \\
$v_\text{eq}$ [\kms]\tablefootmark{d}          & -                         & $240_{-140}^{+300}$          &  $< 300$                      & $< 400$                 & $120^{+30}_{-30}$          & -                       & $180^{+100}_{-80}$           & -                      & $160_{-30}^{+70}$          & -                         & $190^{+40}_{-60}$ \\
$M_{V,\text{John}}$ [mag]                     & $-4.4^{+0.2}_{-0.2}$      & $-3.6^{+0.5}_{-0.5}$  &  $-7.1^{+0.3}_{-0.3}$        & $-6.8^{+0.4}_{-0.4}$   & $-6.7^{+0.3}_{-0.3}$     & $-5.15^{+0.3}_{-0.3}$ & $-6.65^{+0.3}_{-0.3}$  & $-4.4^{+0.3}_{-0.3}$    & $-5.7^{+0.3}_{-0.3}$  & $-4.9^{+0.3}_{-0.3}$    & $-5.9^{+0.3}_{-0.3}$ \\
$X_{\rm H}$ (mass fr.)\tablefootmark{e}       &  0.25$^{+0.05}_{-0.05}$    &  0.73                &  0.25$^{+0.05}_{-0.05}$                &  0.25$^{+0.20}_{-0.20}$           &  0.73               &  0.4$^{+0.1}_{-0.1}$          &  0.73                &  0.15$^{+0.05}_{-0.05}$          &  0.73               &  $0^{+0.15}$                       &  0.73\\
$X_{\rm C}/10^{-5}$ (mass fr.)\tablefootmark{e}     &  $2^{+1}_{-1}$          &  $21 $ & $3^{+1}_{-1}$        & $3^{+2}_{-2}$  & $21$ &  $3^{+1}_{-1}$   & $21$   & $3^{+1}_{-1}$  & $21$   & $0.30^{+0.05}_{-0.05}\cdot10^5$                & $21$              \\
$X_{\rm N}/10^{-3}$ (mass fr.)\tablefootmark{e}   &  $4^{+2}_{-2}$            &  $0.03\,(|\,1\tablefootmark{\text{f}}) $   & $2.5^{+1}_{-1}$  & $2.5^{+1}_{-1}$ & $0.03$   & $3^{+1}_{-1}$   & $0.03\,(|\,0.3\tablefootmark{\text{f}})$    & $2^{+1}_{-1}$  & $0.03\,(|\,0.3\tablefootmark{\text{f}})$   &  0                          &0.03 \\
$X_{\rm O}/10^{-5}$ (mass fr.)\tablefootmark{e}    &  $5.5 $                   &  $110 $ & $5.5 $            & $5.5$      & $110$ & $5.5 $   & $110 $  & 5.5     & 110           &  $0.3^{+0.1}_{-0.1}\cdot10^5$                  & $110 $\\
$E_{B-V}$ [mag]                                &  \multicolumn{2}{c|}{0.18$^{+0.01}_{-0.01}$} &  \multicolumn{3}{c|}{0.08$^{+0.02}_{-0.02}$} &  \multicolumn{2}{c|}{0.065$^{+0.01}_{-0.01}$}  &  \multicolumn{2}{c|}{0.08$^{+0.01}_{-0.01}$}  &  \multicolumn{2}{c}{0.07$^{+0.01}_{-0.01}$} \\
$A_V$ [mag]                                   & \multicolumn{2}{c|}{0.56$^{+0.03}_{-0.01}$} & \multicolumn{3}{c|}{0.25$^{+0.06}_{-0.06}$} & \multicolumn{2}{c|}{0.20$^{+0.03}_{-0.03}$} & \multicolumn{2}{c|}{0.25$^{+0.03}_{-0.03}$}  & \multicolumn{2}{c}{0.22$^{+0.03}_{-0.03}$} \\
$M_{\rm H-b}$ [$M_\odot$]\tablefootmark{g}     & $46^{+7}_{-6}$            & -                     & $83^{+20}_{-16}$             & $73^{+42}_{-25}$          & -                            & $101^{+30}_{-23}$    & -                       &  $51^{+13}_{-10}$           &  -                      &  -                         &  -        \\  
$M_{\rm He-b}$ [$M_\odot$]\tablefootmark{g}     & $29^{+2}_{-2}$            & -                    & $54^{+9}_{-9}$                & $47^{+13}_{-10}$           & -                           & $49^{+9}_{-7}$        & -                       &  $37^{+6}_{-5}$            &  -                      &  $40^{+7}_{-6}$              &  -        \\          
$M_{\rm g}$ [$M_\odot$]                       & -                         & $13^{+70}_{-10}$      & -                             & -                        & $34^{+64}_{-22}$             & -                       & $54^{+97}_{-34}$       &  -                       & $30^{+55}_{-19}$          &                             &  $70^{+210}_{-52}$       \\                                                                                                                                                 
$M_{\rm orb}$ [$M_\odot$]\tablefootmark{h}     & $20^{+80}_{-15}$          & $20^{+20}_{-5}$      & $61^{+10}_{-10}$            & $66^{+10}_{-10}$            & -                         & $9^{+7}_{-3}$      & $41^{+29}_{-16}$        &  $23^{+13}_{-5}$            & $44^{+26}_{-9}$     & $19^{+3}_{-8}$                 &  $61^{+14}_{-25}$  \\                                                                                                                                                                                                                                                                                                                                                                                    
$R_\text{RL}[R_\odot]$\tablefootmark{i}     &  $25^{+29}_{-6}$            & $25^{+11}_{-6}$       &  $58^{+4}_{-4}$             &$60^{+4}_{-4}$                      & -                   & $14^{+3}_{-2}$         & $28^{+9}_{-7}$        & $40^{+8}_{-3}$           & $54^{+13}_{-6}$                      & $33^{+3}_{-5}$              & $57^{+7}_{-12}$ \\
\hline
\end{tabular}
\tablefoot{ All entries but spectral types, $M_\text{orb}$, and $R_\text{RL}$ are derived in this study unless otherwise stated.  Values without errors are adopted.   \\ 
\tablefoottext{a}{References as in Table.\,\ref{tab:overview}}
\tablefoottext{b}{ Fixed for WR components using $M_\text{orb}$ and $R_*$ (see Sect.\,\ref{subsec:method})}
\tablefoottext{c}{ unconstrained entries adopted from \citet{Vink2000}.}
\tablefoottext{d}{ Equatorial rotation velocity calculated assuming alignment of the orbital and rotational axes.} 
\tablefoottext{e}{Entries without errors are fixed to typical SMC abundances (see Sect.\,\ref{subsec:assumptions}) }
\tablefoottext{f}{Alternative values in parentheses obtained when assuming the N\,{\sc iii}\,$\lambda 4640$ emission originates in the 
 O component (see Appendix \ref{sec:comments})}
\tablefoottext{g}{ Obtained from MLRs by \citet{Graefener2011}  (see Sect.\,\ref{subsec:masses})}
\tablefoottext{h}{ Based on orbital parameters given in Table\,\ref{tab:overview} }
\tablefoottext{i}{Calculated via the Eggleton approximation \citep{Eggleton1983} assuming the orbital parameters given in Table\,\ref{tab:overview}}
}
\end{center}
\end{table*}

\section{Results}
\label{sec:results}

Table\,\ref{tab:stellarpar} summarizes the stellar parameters derived for the components of the five systems analyzed. 
The spectral fits are available in Appendix\,\ref{sec:specfits} (Figs.\,\ref{fig:AB3} to \ref{fig:AB8}).
The Table also includes the temperatures and radii at $\tau_\text {Ross} = 2/3$,  H, C, N, O abundances, 
Johnson $V$ magnitudes, projected and equatorial rotation velocities $v \sin i$ and $v_\text{eq}$, 
total reddenings $E_{B - V}$ and extinctions $A_V$, and Roche lobe radii $R_\text{RL}$, 
calculated from the orbital masses using the Eggleton approximation \citep{Eggleton1983}.
We also give several types of stellar masses:
 $M_\text{H-b}$ and $M_\text{He-b}$ are derived for WR stars from MLRs calculated by  \citet{Graefener2011}
for chemically homogeneous core H- and He-burning stars, respectively (see Sect.\,\ref{subsec:masses}).
$M_{\rm g} = G^{-1}\,g_*\,R_*^2$ is inferred from the derived surface gravity.
Finally, $M_{\rm orb}$ denotes the orbital masses, calculated from the orbital parameters given in Table\,\ref{tab:overview}. 

Uncertainties for the fundamental stellar parameters are estimated
by examining the sensitivity of the fits to changes in the corresponding parameters. 
These include errors 
on $T_*\,, \log g_*\,, \log L, \log R_\text{t}, v_\infty, D, E_{\rm B-V}, v \sin i$ and abundances. 
Error propagation is used for the remaining parameters. Errors on $\dot{M}$ include only errors on $R_\text{t}$. Errors on $M_\text{orb}$ are 
dominated by errors on $i$ (cf.\ Table\,\ref{tab:overview}), except for AB\,5, where the uncertainty on the orbital solution 
dominates \citep{Koenigsberger2014}.

\section{Discussion}
\label{sec:disc}

\subsection{Comparison with mass-luminosity relations}
\label{subsec:masses}

The surface of WR stars remains hidden behind 
their stellar winds, rendering a determination of their masses via photospheric absorption lines or astroseismological methods
difficult.
The only method to estimate the masses of single WR stars is by using mass-luminosity 
relations \citep[MLRs, e.g.][]{Langer1989, Graefener2011}. Clearly, these relations need to be calibrated with 
model-independent methods for measuring stellar masses. 
Binary systems offer the most reliable method to "weigh" stars 
using simple Newtonian dynamics, given  that the required observables 
($K_1, K_2, P, i$, and $e$) are known.

We now compare  theoretical MLRs published by \citet{Graefener2011} to the luminosities and orbital masses inferred for the WR primaries in our 
sample\footnote{We note that while these relations 
were calculated at solar metallicity, the 
influence of the metallicity is negligible (G. Gr\"afener, priv.\ com.)}.
The relations are calculated for the simplified case of chemically 
homogeneous stars. In these relations, $\log L$ is given as 
a second order polynomial in $\log M$ (see Eqs.\,9 and 10 in \citealt{Graefener2011}), where the coefficients depend 
on the hydrogen mass fraction $X_\text{H}$. In these relations, if $X_\text{H} > 0$, the star is assumed 
to be core H-burning. Otherwise, it is assumed to be core He-burning. 

For a star with a given luminosity, the largest possible mass predicted by theory
is obtained for chemically homogeneous, hydrogen burning stars (Eq.\,11 in \citealt{Graefener2011}). 
Lower masses are obtained if the star has a He-burning core.
\citet{Graefener2011} argue that 
the MLR derived for pure helium stars (Eq.\,13 in \citealt{Graefener2011}) should give a good approximation 
for the masses of evolved, He-burning WR stars. However, if the contribution of shell H-burning to the luminosity is 
significant, a given luminosity can be supported by a yet smaller mass. 
The most strict lower bound on the mass at a given luminosity is given by the classical 
Eddington limit calculated for a fully ionized atmosphere (including only electron scattering).

\begin{figure}[!htb]
\centering
  \includegraphics[width=\columnwidth]{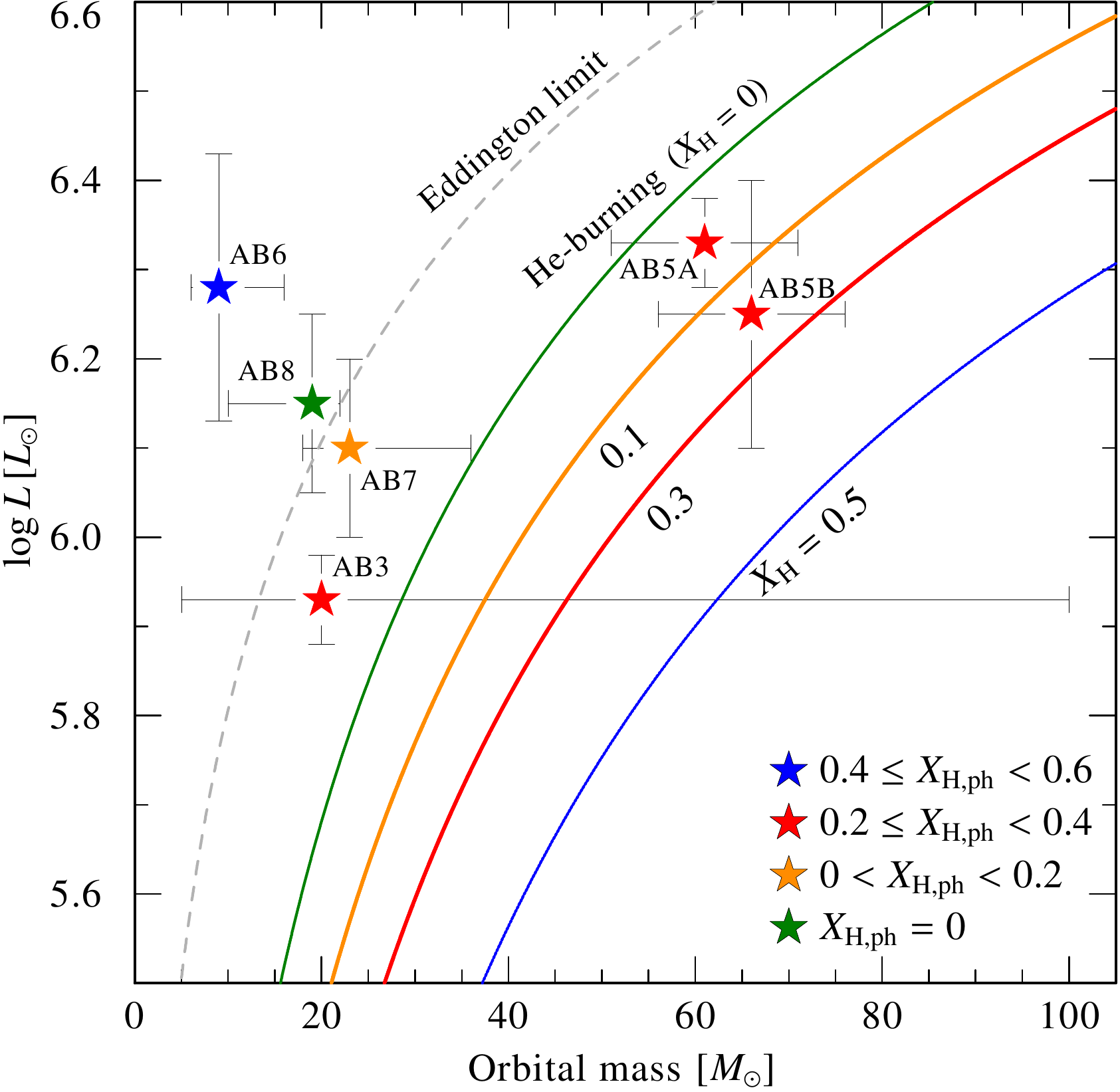}
  \caption{Positions of the WR components on a $M - \log L$ diagram (symbols) compared to 
  MLRs calculated for homogeneous stars \citep{Graefener2011}, depicted by solid curves. 
  The colors correspond to the hydrogen content (see legend). The Eddington limit 
  calculated for a fully ionized helium atmosphere is also plotted (gray dashed line).}
\label{fig:orbmass}
\end{figure}

Fig.\,\ref{fig:orbmass} compares the theoretical predictions of the MLRs to the empirically derived $(M_\text{orb}, \log L)$ coordinates for the  WR companions. 
The Eddington limit, calculated for a fully ionized helium atmosphere, is also plotted. \citet{Langer1989} also provides calculations 
for homogeneous stars with a vanishingly small helium abundance, which may be more suitable for the WO component in AB\,8. Since these calculations 
predict a very similar relation to the MLR calculated for pure He-stars (the latter predicting slightly lower luminosities for a given 
mass), we omit the corresponding MLR from Fig.\,\ref{fig:orbmass} for clarity.

Both massive 
WR components of AB\,5 are located between MLRs calculated with $X_\text{H,core} = 0$ and $0.3$. This suggests that these 
stars may still be core H-burning, although both could coincide with the relation for He-burning stars within errors.
Since the WR components of AB\,3, 7, 6, and 8 are of early spectral type, which 
are understood to be core He-burning stars, we can expect them to lie on the MLR for pure He-stars, or,
if a significant fraction of the luminosity originates in shell burning, between this relation and the Eddington limit\footnote{We note that, at a given temperature, 
the spectral types of WR stars in the SMC 
tend to appear ``earlier'' than their Galactic counterparts \cite[e.g.][]{Crowther2006}. Seemingly early type stars in the 
SMC could therefore still be core H-burning.}. 
As is apparent in Fig.\,\ref{fig:orbmass},
these stars do lie above the pure He-star MLR,
suggesting that they are indeed core helium burning. The WR component in AB\,3 is poorly constrained due to the large error on $M_\text{orb}$.
Since all analyzed WN components 
show signatures of hydrogen in their atmospheres, this suggests
that the majority of WR stars in our sample are not chemically homogeneous. Taken at face value, their offsets 
from the MLR calculated for pure helium stars suggests the presence of shell H-burning (or shell He-burning in the case of AB\,8).

While the WR component of AB\,7 is located below the Eddington limit, that of AB\,8 (WO) slightly exceeds it. 
The internal structure of WO stars is poorly understood and hard to model (N.\ Langer, priv.\ com.). 
Regardless, it should not be possible for the star to exceed the Eddington limit, unless 
significant departures from spherical symmetry occur already in the stellar interior \citep[e.g.][]{Shaviv2000}.
However, considering the given errors on $\log L$ and $M_\text{orb}$, there is no clear discrepancy in the case of AB\,8.

The only star in the sample for which a clear inconsistency is obtained is the WR component of the shortest-period binary in our 
sample, AB\,6. The star clearly exceeds its Eddington limit, which immediately implies that the derived luminosity and/or the orbital mass 
are incorrect. Furthermore, we find a significant amount of hydrogen ($\approx 40\%$) in its atmosphere, 
which is not expected for highly evolved, He-burning stars.
In Sect.\,\ref{subsec:WR6}, we discuss possible reasons for this discrepancy, and argue that the orbital mass derived for the WR companion is most likely 
wrong. It is therefore omitted when considering the evolutionary status of the system in the next sections.

\subsection{Evolutionary status: avoiding mass-transfer due to homogeneous evolution}
\label{subsec:sinevo}

Given the short orbital periods of our objects,
it seems likely that the primaries underwent a RLOF phase before becoming WR stars. 
However, rapid initial equatorial rotation in excess of $400\,$\kms\, \citep{Heger2000I, Brott2011} 
may lead to quasi-chemically homogeneous evolution  (QCHE). 
A star experiencing QCHE  maintains higher effective temperatures and thus much smaller radii throughout its evolution, and 
may therefore avoid overfilling its Roche lobe during the pre-WR phase.

In paper I, we argued that the properties of the single SMC WR stars are compatible with QCHE. 
Yet tidal forces in binaries act to synchronize the axial rotation of the components with the orbital period \citep{Zahn1977}. 
In binaries of extremely short periods ($\lesssim 2\,$d), synchronization can maintain, or even enforce, near-critical rotation of the 
components \citep[e.g.][]{DeMink2009, Song2016}. However, 
none of the binaries in our sample portray such short orbital periods, and we find no evidence for a significant increase of the orbital period throughout their
evolution (see Sect.\,\ref{subsec:evolution}). In our sample, tidal interactions are rather expected to have slowed down the stellar rotation.
For example, for an O star of $10-20\,R_\odot$ (a typical main sequence WR progenitor), 
synchronization with the period of AB\,5 (19.3\,d) would imply a rotational velocity of only 50-100\,\kms. Even 
for AB\,6, the shortest period binary in our sample (6.5\,d), synchronization implies $150-300$\,\kms, which is insufficient 
to induce QCHE.

\renewcommand{\arraystretch}{1.1}

\begin{table*}[!htb]
\scriptsize
\small
\caption{Initial parameters and ages derived from single-star tracks assuming inhomogeneous evolution for the primary} 
\label{tab:evocompsinnon}
\begin{center}
\begin{tabular}{l | cc  cc cc  cc  cc }
\hline
                                    & \multicolumn{10}{c}{Inhomogeneous primaries, single-star tracks} \\    
\hline \rule{0mm}{3mm}     
       SMC\,AB\,                                 & \multicolumn{2}{c}{3}        & \multicolumn{2}{c}{5}        &\multicolumn{2}{c}{6}       &\multicolumn{2}{c}{7}       & \multicolumn{2}{c}{8}       \\
\hline \rule{0mm}{3mm}                                                                                                                                                              
$M_\text{1,\,i}[M_\odot]$\tablefootmark{a}        &   \multicolumn{2}{c}{50}     &  \multicolumn{2}{c}{100}       &   \multicolumn{2}{c}{80}   &     \multicolumn{2}{c}{70}     &  \multicolumn{2}{c}{80}        \\
Age\,[Myr]\tablefootmark{a}                      &   \multicolumn{2}{c}{4.6}    &   \multicolumn{2}{c}{3.0}     &    \multicolumn{2}{c}{3.4} &  \multicolumn{2}{c}{3.7}          & \multicolumn{2}{c}{3.6}       \\ 
$R_\text{max, 1}[R_\odot]$\tablefootmark{a}                           &   \multicolumn{2}{c}{1800}     &  \multicolumn{2}{c}{1200}  &   \multicolumn{2}{c}{2000}   &     \multicolumn{2}{c}{2100} &  \multicolumn{2}{c}{2000}        \\
$M_\text{i,\,2}[M_\odot]$\tablefootmark{b}                       &   \multicolumn{2}{c}{15}     &   \multicolumn{2}{c}{100}      &  \multicolumn{2}{c}{46}    &  \multicolumn{2}{c}{35}            &  \multicolumn{2}{c}{50}     \\
$v_\text{rot,\,i,\,2}$[\kms]\tablefootmark{b}       &   \multicolumn{2}{c}{230}    &   \multicolumn{2}{c}{-}  &  \multicolumn{2}{c}{170}          &  \multicolumn{2}{c}{160}       &  \multicolumn{2}{c}{130}     \\
\hline \rule{0mm}{3mm}                                                                                   
$\log \left(T_1^\text{E} / T_1^\text{O}\right)$        &   -0.01        & (0.03)           &  -0.01        & (0.05)    &  -0.01                  & (0.06)         &  -0.02  &  (0.04)          &  0.09  & (0.15)   \\
$\log \left(L_1^\text{E} / L_1^\text{O}\right)$        &   -0.08        & (0.08)           &  -0.04        & (0.13)    & -0.16                   &   (0.15)       &  -0.07  &  (0.12)          &  -0.09 & (0.13)    \\
$M_1^\text{E} - M_1^\text{O}$                     &    2           & (50 )            &  -13          & (14)          &  -                        & -            &  8      & (14)            &  8    &  (10 )       \\
$X_\text{H,1}^\text{E} - X_\text{H,1}^\text{O}$   &   -0.03        & (0.05)           & \underline{-0.11}   & (0.05)      &  \underline{-0.25}  & (0.1)          &  -0.03  &  (0.05)          &  0.00  & (0.05) \\ 
\hline
\end{tabular}
\tablefoot{
The upper part of the table gives the initial parameters and age defining the best-fitting single star tracks assuming 
a non-homogeneous evolution, as derived by the BPASS and BONNSAI tools, as well as the maximum radius reached by the primary throughout the evolution, $R_\text{max, 1}$. 
The lower part gives $\text{E}_n - \text{O}_n$ as obtained from the best-fitting BPASS track, along 
with corresponding $\sigma_n$ values in parentheses (see Sect.\,\ref{subsec:sinevo}). Underlined values denote deviations which exceed $2\sigma$.\\
\tablefoottext{a}{Obtained from the BPASS stellar evolution code}
\tablefoottext{b}{Obtained from the BONNSAI stellar evolution tool, except for AB\,5, where the BPASS code was used (see text) }
}
\end{center}
\end{table*}

\begin{table*}[!htb]
\scriptsize
\small
\caption{Initial parameters and ages derived from single-star tracks assuming homogeneous evolution for the primary}
\label{tab:evocompsinhomo}
\begin{center}
\begin{tabular}{l | cc  cc cc  cc  cc }
\hline      
                                    & \multicolumn{10}{c}{Homogeneous primaries, single-star tracks} \\    
 \hline \rule{0mm}{3mm}     
       SMC\,AB\,                                 & \multicolumn{2}{c}{3}                & \multicolumn{2}{c}{5}                     &\multicolumn{2}{c}{6}       &\multicolumn{2}{c}{7}       & \multicolumn{2}{c}{8}       \\                                                                                                                                                                                             
\hline \rule{0mm}{3mm}                                                                                                              
$M_\text{i,\,1}[M_\odot]$\tablefootmark{a}        &   \multicolumn{2}{c}{50}             &  \multicolumn{2}{c}{70}                    &   \multicolumn{2}{c}{100}       &     \multicolumn{2}{c}{50}                   &  \multicolumn{2}{c}{70}        \\
Age\,[Myr]\tablefootmark{a}                      &   \multicolumn{2}{c}{4.5}            &   \multicolumn{2}{c}{3.4}                  &    \multicolumn{2}{c}{2.2}     &  \multicolumn{2}{c}{5.4}       & \multicolumn{2}{c}{4.6}       \\ 
$R_\text{1,\,max}[R_\odot]$\tablefootmark{a}     &   \multicolumn{2}{c}{10}             &  \multicolumn{2}{c}{13}               &   \multicolumn{2}{c}{19}   &     \multicolumn{2}{c}{10} &  \multicolumn{2}{c}{13}        \\
$M_\text{i,\,2}$\tablefootmark{b}                 &   \multicolumn{2}{c}{15}             &   \multicolumn{2}{c}{70}                  & \multicolumn{2}{c}{55} & \multicolumn{2}{c}{No solution}    & \multicolumn{2}{c}{40}       \\      
$v_\text{rot\,i,\,2}$\tablefootmark{b}            &   \multicolumn{2}{c}{230}            &   \multicolumn{2}{c}{-}       & \multicolumn{2}{c}{170} & \multicolumn{2}{c}{ No solution}       & \multicolumn{2}{c}{410} \\
\hline \rule{0mm}{3mm}                                                                                   
$\log \left(T_1^\text{E} / T_1^\text{O}\right)$        &   \underline{-0.12}  & (0.03)      &  \underline{0.13}  & (0.05)              &  \underline{-0.13}  & (0.06)       &   0.01                        &  (0.06)               &  0.09  & (0.15)   \\
$\log \left(L_1^\text{E} / L_1^\text{O}\right)$        &   0.06                & (0.08)      &  -0.13            & (0.13)              & 0.05                &   (0.16)  &   -0.08                           &  (0.12)                   &  -0.08 & (0.13)    \\
$M_1^\text{E} - M_1^\text{O}$                         &    28                  & (80 )       &  6               & (11)                 &  -                  & -              &         10                & (10 )      &  8     & (10 )       \\
$X_\text{H,1}^\text{E} - X_\text{H,1}^\text{O}$       &   -0.03               & (0.05)      & -0.01              & (0.05)               &  -0.03             & (0.1)             &   \underline{-0.15}       &  (0.05)      &  0.00  & (0.05) \\  
\hline
\end{tabular}
\tablefoot{
Same as Table\,\ref{tab:evocompsinnon}, but assuming homogeneous evolution (for footnotes, see Table\,\ref{tab:evocompsinnon}).
}
\end{center}
\end{table*}

Synchronization timescales $\tau_\text{sync}$ involve much uncertain physics. \citet{Hurley2002} give some estimates for binary stars with a mass ratio $q = 1$ and 
for initial masses up to $M_\text{i} = 10\,M_\odot$.
They show that 
for stars with $M_\text{i} > 1\,M_\odot$ (i.e.\ stars with radiative envelopes), 
separations of the order of $10\,R_*$ ensure a synchronization timescale $\tau_\text{syn}$ which is smaller 
than the main sequence timescale $\tau_\text{MS}$,  with the ratio $\tau_\text{syn} / \tau_\text{MS}$ virtually independent of the mass.
Since the systems analyzed here are characterized by separations of a few $R_*$,
tidal interactions are expected to have greatly lowered the initial rotation rates.

To test whether single-star evolutionary tracks can explain the observed properties of our objects, 
we compare the primaries' $T_*, L, M_\text{orb}$, and $X_\text{H}$ with
evolutionary tracks for single stars with initial masses between $20\,M_\odot$ and $120\,M_\odot$ calculated 
at a metallicity of $Z = 0.004$
with the  BPASS\footnote{bpass.auckland.ac.nz} 
(Binary Population and Spectral Synthesis) stellar evolution code \citep[][Eldridge et al.\ in prep.]{Eldridge2008}, 
which can treat both 
single and binary stars. 
We use two sets 
of tracks, one calculated assuming no chemical mixing, and the other calculated assuming a homogeneous evolution \citep{Eldridge2011, Eldridge2012}. 
For each WR star in our sample, we look for a track defined by $M_\text{i}$, and for an age $t$, 
which reproduce the observed quantities $T_*, L, M_\text{orb}$ and $X_\text{H}$ as 
good as possible, in the sense of minimizing the sum

\begin{equation}
 \chi^2\left(M_\text{i}, t\right) =\sum_{n=1}^{4} 
 \left(\frac{\text{O}_n - \text{E}_n\left(M_\text{i}, t\right)}{\sigma_n}\right)^2,
\label{eq:summinsin}
\end{equation}
where $\text{O}_n \in \left\{ \log T_*, \log L, M_\text{orb}, X_\text{H} \right\}$ are the inferred values for the
considered observables, and $\text{E}_n\left(M_\text{i}, t\right)$ are the corresponding 
predictions of the evolutionary track defined by the initial mass $M_\text{i}$ at age $t$.  Only in the case of AB\,6, we 
ignore the WR component's orbital mass because of its clear inconsistency with the derived stellar luminosity (see Sects.\,\ref{subsec:masses} and \ref{subsec:WR6}).
Since the tracks evolve non linearly, we avoid interpolation over the grid. 
 Instead, we define $\sigma_n = \sqrt{\Delta_n^2 + \delta_n^2}$, where $\Delta_n$ is half the $n$'th parameter's 
grid spacing, and $\delta_n$ is the corresponding error given in Table\,\ref{tab:stellarpar}.  In the case of 
asymmetrical errors in Table\,\ref{tab:stellarpar}, we assign $\delta_n$
according to whether $\text{O}_n> \text{E}_n $ or $\text{O}_n  < \text{E}_n $.
By minimizing $\chi^2$, we  infer initial masses and ages for the primaries 
in the cases of no mixing and homogeneous evolution. Conservative uncertainties on the ages are constrained from the adjacent tracks
in the vicinity of the solution (typically $0.2\,$Myr).

\begin{figure*}
\centering
\begin{subfigure}{\columnwidth}
  \centering
  \includegraphics[width=0.95\linewidth]{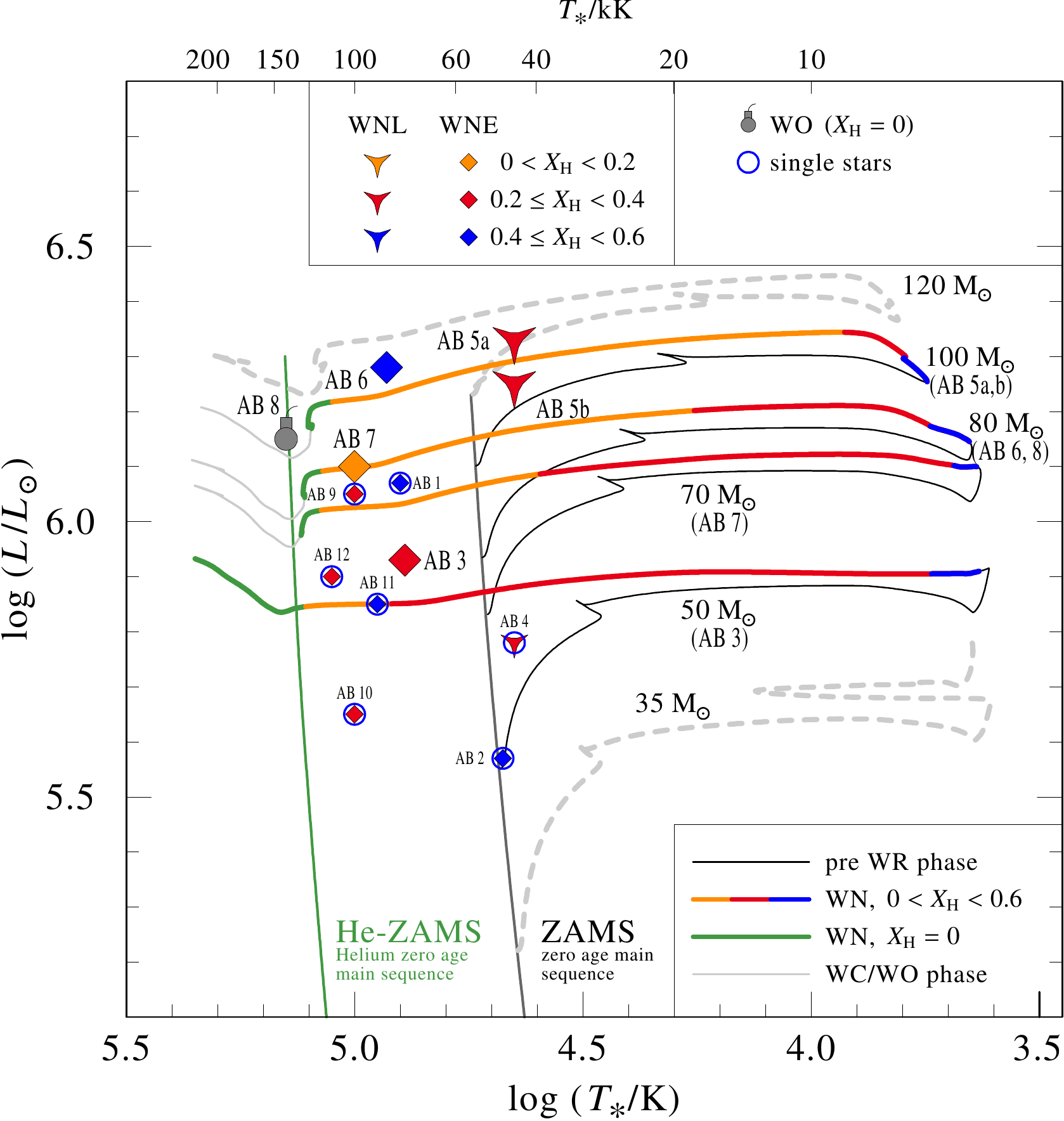}
  \label{fig:sub1}
\end{subfigure}%
\begin{subfigure}{\columnwidth}
  \centering
  \includegraphics[width=0.95\linewidth]{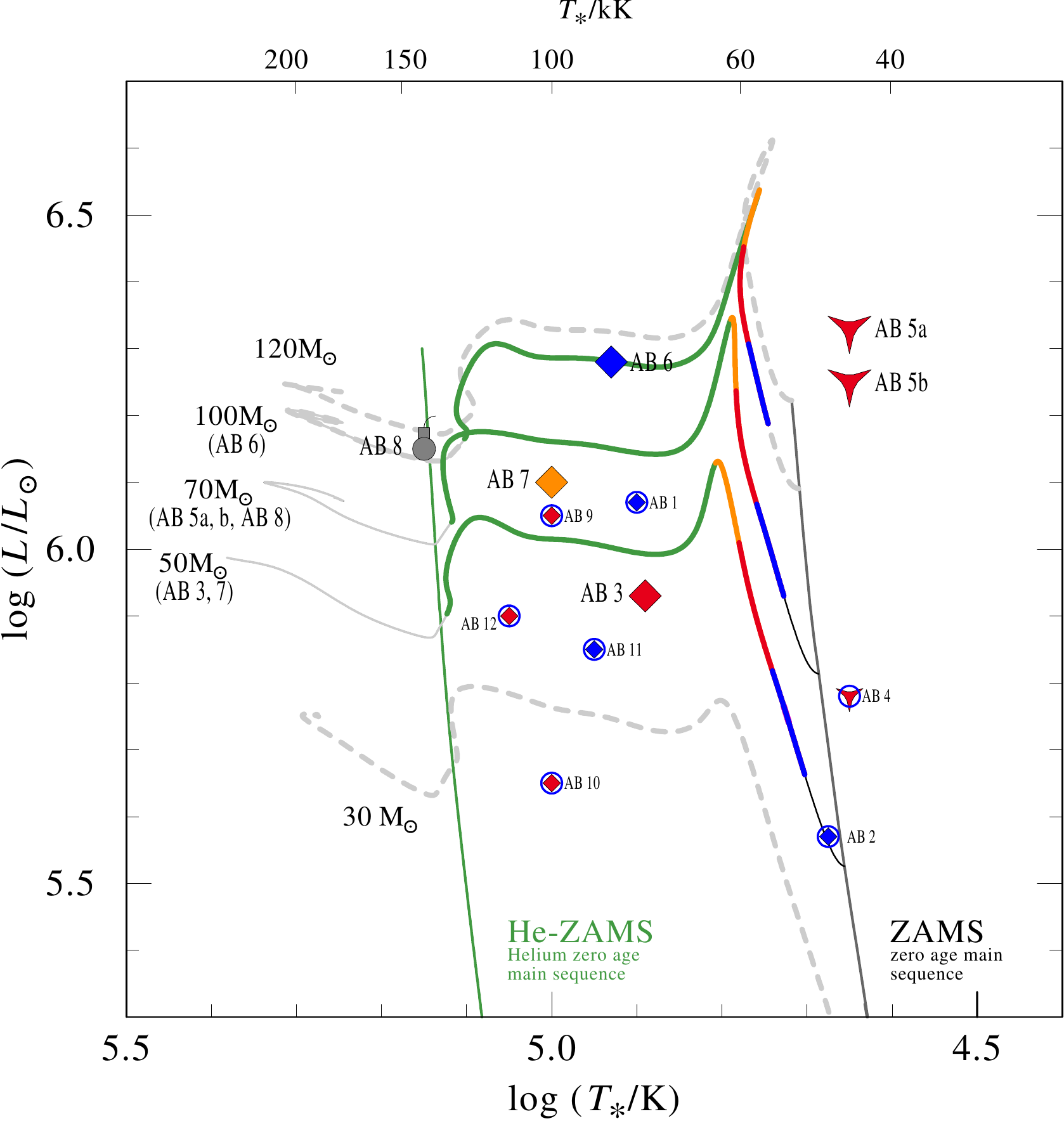}
  \label{fig:sub2}
\end{subfigure}
\caption{HRD positions of the all known single (paper I) and binary (this paper) SMC WR stars. The big symbols 
    correspond to the binaries analyzed here, while circled smaller symbols correspond to putatively single stars.  
      Plotted are evolutionary tracks \citep{Eldridge2008} calculated for single stars 
      at a metallicity of $Z = 0.004$ assuming no mixing (left panel) and chemically homogeneous evolution (right panel). 
      The colors and symbols 
      code the hydrogen abundance and WR type, as described in the legend.  The WR phase is defined at $X_\text{H} < 0.6$.
      Tracks which correspond to 
      the solutions found are shown in color.}
\label{fig:HRDsin}
\end{figure*}

As a second step, we test whether the ages derived
are consistent with the current evolutionary status of the secondary,  assuming still that no interaction has occurred between the companions.
For this purpose, we use the BONNSAI\footnote{The BONNSAI web-service is available at www.astro.uni-bonn.de/stars/bonnsai} Bayesian statistics tool \citep{Schneider2014}. 
The tool interpolates over detailed evolutionary tracks calculated by \citet{Brott2011}
for stars of initial masses up to $60\,M_\odot$ and over a wide range of initial rotation velocities $v_\text{rot, i}$.
Using $T_*, L, M_\text{orb}, v \sin i$ derived for the secondary, as well as the age derived for the primary (along 
with their corresponding errors), the algorithm tests whether a model exists which can reproduce the secondary's properties at the derived age
at a $5\%$ significance level. In case no such model was found, we verified this is not a consequence of the uncertain evolution of 
$v_\text{eq}$ by lifting the $v \sin i$ constraint. For AB\,5, we use the BPASS tracks to check consistency with the secondary, since its mass is not covered by 
the BONNSAI tool. 

Tables\,\ref{tab:evocompsinnon} and \ref{tab:evocompsinhomo} show the initial masses $M_\text{i,1}$ and ages inferred for the primaries in the cases of
inhomogeneous/homogeneous evolution.  The Tables also give the maximum radius reached by the primary along the best-fitting track, $R_\text{max, 1}$.
If consistent solutions for the secondaries are found by the BONNSAI tool, the secondaries' initial masses $M_\text{i, 2}$ and rotations $v_\text{rot, i}$, 
as obtained from the BONNSAI tool, are given. 
Tables\,\ref{tab:evocompsinnon} and \ref{tab:evocompsinhomo} also gives the differences $\text{O}_n - \text{E}_n$ for each of the primary's parameters, 
where we also include $\sigma_n$ values. 
While all solutions provided by the BONNSAI reproduce the observed properties of the secondaries at a 5\% significance level \citep{Schneider2014}, 
they do not do so equally well.
However, since this is merely a consistency test for single-star evolution, we do not present a detailed description of the BONNSAI fit quality, which 
can be recovered online by the interested reader.

The two panels of Fig.\,\ref{fig:HRDsin} show the positions of the complete SMC WR population in a $\log L - \log T_*$ diagram (HRD), as derived in 
paper I and in this study. 
The left panel includes BPASS tracks for the primaries calculated assuming no mixing, 
while the right panel includes BPASS tracks assuming homogeneous evolution. The obtained solutions are highlighted in color. 
Note that the HRD contains only partial information 
regarding the fit quality (see Tables\,\ref{tab:evocompsinnon} and \ref{tab:evocompsinhomo}). For clarity, we do not include error bars in Fig.\,\ref{fig:HRDsin}.

From this test, it seems that QCHE is not consistent with AB\,3, 6, and 7. For example, the $T_*$ predicted by the track best fitting AB\,3 deviates
by 4$\sigma$ from our measurement. 
For AB\,7, not only the hydrogen content is underpredicted,  but also, the age is not consistent with the secondary's stellar parameters.
 Based on our results, QCHE does not seem consistent with AB\,5 either, since the temperature of the primary is overpredicted by more than $2\sigma$. However,
\citet{Koenigsberger2014} manage to explain the evolutionary status of AB\,5 by assuming non-interacting companions experiencing QCHE. This discrepancy 
occurs because of the lower value inferred for $T_*$ in this work compared to that used by \citet{Koenigsberger2014}. Indeed, WWC in AB\,5 may be responsible 
for a systematic uncertainty on $T_*$ (see Appendix\,\ref{sec:comments}).
Of all systems, only AB\,8 is compatible with homogeneous evolution.

The tracks which do not include mixing generally show a better agreement.
However,  the primary stars in this set of evolutionary tracks reach radii which greatly exceed their Roche lobe radii (cf.\, Table\,\ref{tab:evocompsinnon}). If the systems did not undergo QCHE, there is little doubt that their companions have interacted via mass-transfer. In the next section, 
we account for this effect by considering binary evolution models.

\subsection{Evolutionary status: assessing binary effects}
\label{subsec:evolution}

\begin{figure*}
\centering
\begin{subfigure}{\columnwidth}
  \centering
  \includegraphics[width=0.95\linewidth]{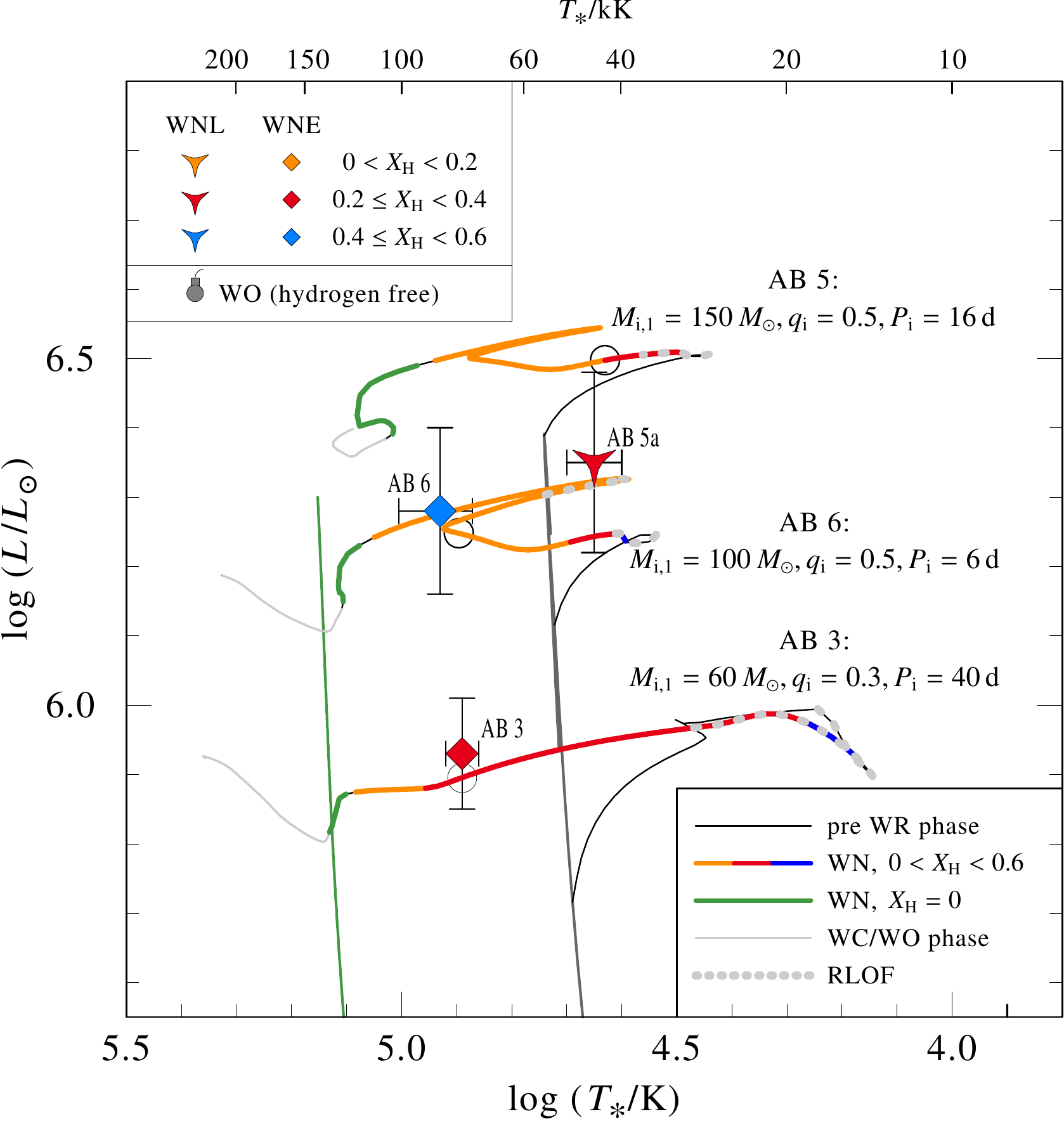}
  \label{fig:sub2}
\end{subfigure}
\begin{subfigure}{\columnwidth}
  \centering
  \includegraphics[width=0.95\linewidth]{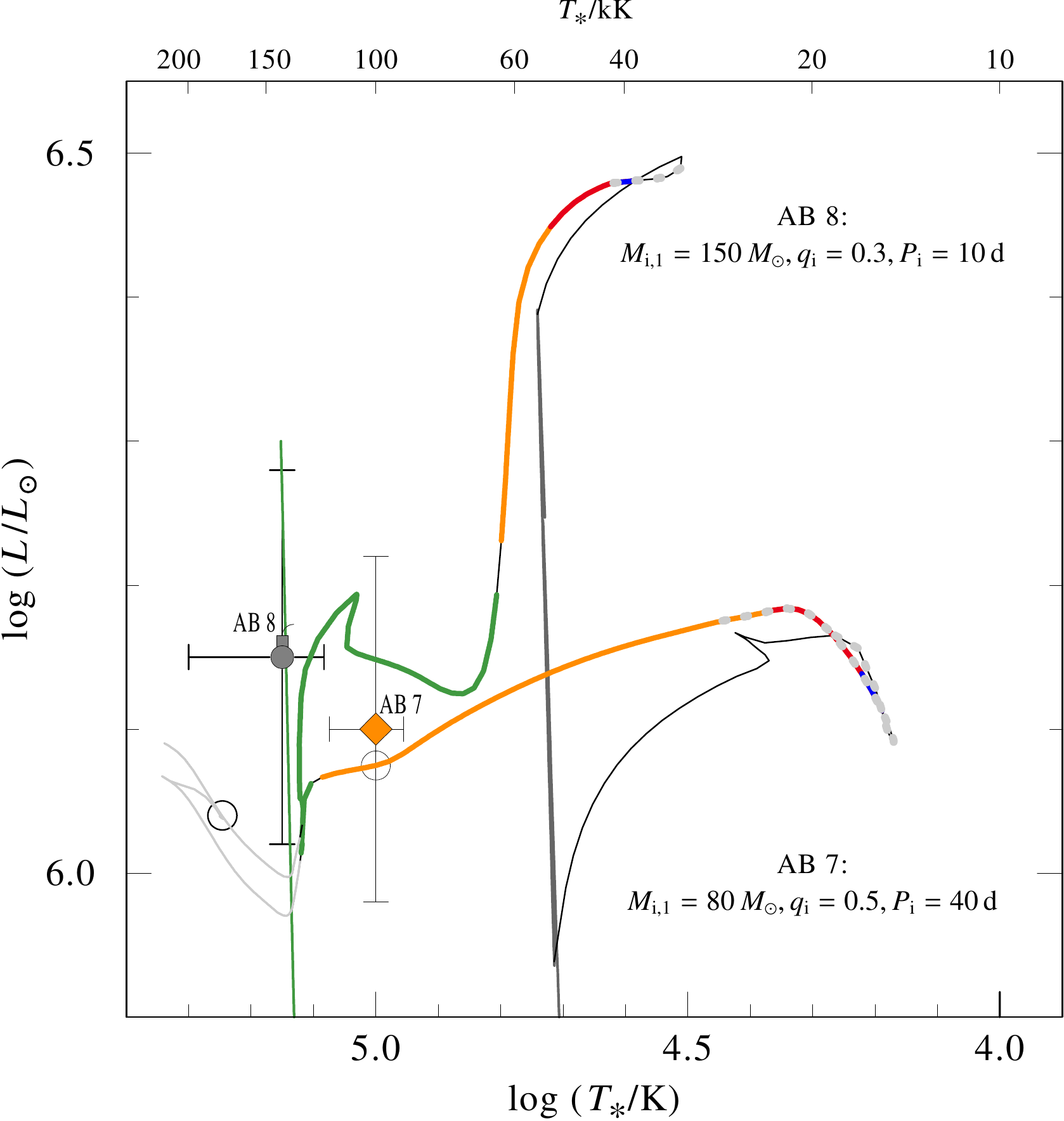}
  \label{fig:sub1}
\end{subfigure}%
\caption{Derived HRD positions of the WR components compared to 
  the  binary evolution tracks which best reproduce the set of eight observables (see text). The tracks shown correspond to the primary component. 
  Colors are as in Fig.\,\ref{fig:HRDsin}. The circles  denote the best-fitting 
  position (ages) along the tracks. RLOF phases 
  are marked with dotted gray lines.}
\label{fig:HRDbin}
\end{figure*}

We would now like to compare the HRD positions of the binary systems to evolutionary tracks which account for binary interaction. 
modeling the evolution of binaries is difficult, because on top of the complex physics
involved in the evolution of single stars, the effects of tidal interaction and mass-transfer have to be accounted for.
While codes exist which account for these effects simultaneously \citep[e.g.][]{Cantiello2007}, there are no corresponding 
grids of tracks available. Here, 
we make use of evolutionary tracks calculated with Version 2.0
of the BPASS code, which accounts for mass-transfer. 
The tracks do not include rotationally induced mixing or tidal interaction. However,
as discussed in Sect.\,\ref{subsec:sinevo}, mixing should be negligible for the majority of our objects.  If mixing does become important in a binary system, 
its components will likely avoid RLOF, as can be inferred from the small radii maintained by the homogeneous models (cf.\, Table\,\ref{tab:evocompsinhomo}). In this 
case, the solutions found from the single-star, chemically homogeneous evolutionary tracks should be adequate to describe the system.

Each binary track is defined by a set of three parameters: the initial mass of the primary $M_{\rm i, 1}$, the initial orbital
period $P_\text{i}$, and the  mass ratio $q_\text{i} = M_\text{i,2} / M_\text{i,1}$. The tracks  
are calculated at intervals of $0.2$ on $0 < \log P\,[\text{d}] < 4$,  $0.2$ on $0 < q_\text{i} < 0.9$, 
and at unequal intervals of $10-30\,M_\odot$ on $10 < M_\text{i,1} < 150\,M_\odot$.
Again, we use a $\chi^2$ minimization algorithm to find the best-fitting track for each system. However, 
this time we consider eight different observables, thus leading to

\begin{equation}
 \chi^2\left(P_\text{i}, q_\text{i}, M_\text{i,\,1}, t\right) =\sum_{n=1}^{8} 
 \left(\frac{\text{O}_n - \text{E}_n\left(P_\text{i}, q_\text{i}, M_\text{i,1}, t\right)}{\sigma_n}\right)^2,
\label{eq:summin}
\end{equation}
where $\text{O}_n \in \left\{ \log T_1, \log L_1, \log T_2, \log L_2, M_\text{orb,\,1}, M_\text{orb,\,2}, \log P,\right.$
$\left. X_\text{H, 1} \right\}$ are the measured values for the
considered observables, and
$\text{E}_n\left(P_\text{i}, q_\text{i}, M_\text{i,1}, t\right)$ are the corresponding 
predictions of the evolutionary track defined by $P_\text{i}$, 
$q_\text{i}$, and $M_\text{i,1}$ at time $t$. $\sigma_n$ is defined as in Eq.\,\ref{eq:summinsin}.

\renewcommand{\arraystretch}{1.2}

\begin{table*}[!htb]
\scriptsize
\small
\caption{Comparison with the best fitting binary evolutionary tracks} 
\label{tab:evocomp}
\begin{center}
\begin{tabular}{l  cc  cc  cc  cc  cc }
\hline
       SMC\,AB\,                                 & \multicolumn{2}{c}{3}        & \multicolumn{2}{c}{5}   &\multicolumn{2}{c}{6}                 &\multicolumn{2}{c}{7}       & \multicolumn{2}{c}{8}       \\
\hline                                                                                   
$M_\text{i,\,1}[M_\odot]$                           &     \multicolumn{2}{c}{60}  &  \multicolumn{2}{c}{150}  &   \multicolumn{2}{c}{100}         &     \multicolumn{2}{c}{80}   &  \multicolumn{2}{c}{150}        \\
$q_{\rm i}(M_\text{i,\,2}/M_\text{i,\,1})$          &    \multicolumn{2}{c}{0.3} &   \multicolumn{2}{c}{0.5} &   \multicolumn{2}{c}{0.5}        &    \multicolumn{2}{c}{0.5}   &   \multicolumn{2}{c}{0.3}      \\
$P_\text{i}$[d]                                    &    \multicolumn{2}{c}{40}  &   \multicolumn{2}{c}{16}  &   \multicolumn{2}{c}{6}           &     \multicolumn{2}{c}{40}   &   \multicolumn{2}{c}{10}       \\
Age\,[Myr]                                           &   \multicolumn{2}{c}{3.9} &   \multicolumn{2}{c}{2.6}      &    \multicolumn{2}{c}{3.0}   &  \multicolumn{2}{c}{3.4}  & \multicolumn{2}{c}{3.0}       \\ 
\hline                                                                                  
$\log \left(T_1^\text{E} / T_1^\text{O}\right)$        &   0.00   & (0.03)           &  -0.02        & (0.05)          &  -0.01       & (0.06)          &  -0.02  &  (0.08)               &  0.09  & (0.15)   \\
$\log \left(L_1^\text{E} / L_1^\text{O}\right)$        &    -0.03 & (0.08)           &  0.16        & (0.13)          &  -0.03  & (0.16)         &  -0.03  &  (0.12)               &  -0.11 & (0.13)    \\
$\log \left(T_2^\text{E} / T_2^\text{O}\right)$        &   0.08   & (0.09)           &  -0.02       & (0.07)          &  0.04        & (0.08)           &  0.08   &  (0.08)               &  -0.06 & (0.08)         \\
$\log \left(L_2^\text{E} / L_2^\text{O}\right)$        &   0.09   & (0.33)           &  -0.17       & (0.26)          &  -0.16        & (0.28)         &  0.03  &  (0.28)               &  -0.22 & (0.27)         \\
$M_1^\text{E} - M_1^\text{O}$                     &   6      & (50  )           &  26            & (18)          &  -    & -            &  11       & (14  )               &  7     & (10  )       \\
$M_2^\text{E} - M_2^\text{O}$                     &   -2     & (16  )           &  31           &  (18)          &  30                & (30)                     &  -1      & (14  )               &  10    & (20  )     \\
$\log \left(P^\text{E} / P^\text{O}\right)$       &   0.11   & (0.10)           &   -0.07       & (0.10)         &   0.02             & ( 0.1)                  &  0.03   & (0.1 )               &  0.02  & (0.10)   \\
$X_\text{H,1}^\text{E} - X_\text{H,1}^\text{O}$   &   -0.03   & (0.05)           & -0.03         & (0.05)        &  \underline{-0.25} & (0.1)                        &  -0.01   &  (0.05)               &  0.00  & (0.05)    \\
\hline                                                                                                       
\end{tabular}
\tablefoot{
The upper part of the table gives the parameters defining the best fitting evolutionary
track and corresponding ages. The lower part gives $\text{E} - \text{O}$ for each parameter, with corresponding 
$\sigma$ values in parentheses. Underlined values denote deviations which exceed $2\sigma$. }
\end{center}
\end{table*}

The two 
panels in Fig.\,\ref{fig:HRDbin} show the best-fitting evolutionary tracks corresponding to the primary components 
along with their HRD positions.  The circles correspond to the current positions (ages) derived. We stress, however, that the HRD illustrates only three 
of the eight observables which were fit here. 
In Table\,\ref{tab:evocomp}, we give the set of initial parameters defining the best-fitting tracks and ages found for each system, along with the 
differences $\text{O}_n - \text{E}_n$ and the corresponding uncertainties $\sigma_n$. 
Evidently, we manage to find tracks which reproduce the eight observables 
within a 2$\sigma$ level for all systems except AB\,6. An evolutionary scenario which includes mass-transfer 
thus appears to be consistent with AB\,3, 5, 7, and 8, although AB\,8 was also consistent with QCHE.
In Appendix \ref{sec:inditracks}, we give a thorough description of the 
evolution of each system as given by the corresponding best-fitting track.

The solution for AB\,5 
overpredicts the components' masses by almost 2$\sigma$, and is 
generally very sensitive to the weighting of the different observables (e.g.\, small changes in $\sigma_n$). 
However, we believe this is simply a result of the grid spacing (see Appendix \ref{sec:inditracks}). 
A greater challenge lies in explaining the similar hydrogen abundances of the two components
The BPASS code does not follow the hydrogen abundance of the secondary, but 
since the two components were born with quite different masses in the derived solution ($150$ and $75\,M_\odot$), it is unlikely that they would evolve 
to a state of significant 
hydrogen depletion simultaneously. A conceivable resolution within the framework of binary evolution could involve 
the secondary losing much of its hydrogen envelope by undergoing a non-conservative RLOF, 
but this would likely require some fine tuning of the initial conditions.
The QCHE scenario thus appears more natural in the case of AB\,5, as proposed by \citet{Koenigsberger2014}.
However, it is not clear whether this scenario is consistent with the presence of tidal forces in the system. 
We discuss this system thoroughly in Appendix\,\ref{sec:inditracks}.

As for AB\,6,  even when omitting the primary's orbital mass in the fitting procedure, we obtain a $\approx 2.5\sigma$ discrepancy 
in the hydrogen abundance, which is found to be lower in the evolutionary track. However, our tests show that this discrepancy is lifted when using
tracks which assume a lower metallicity ($Z = 0.002$), i.e.\ this is a direct result of the uncertain mass-loss rates during the WR phase. 
Moreover, the BPASS binary models do not evolve the secondary in detail and therefore do not include the
secondary overfilling its Roche lobe to transfer material back to the primary, as was reported for other stars \citep[e.g.][]{Groh2008}. Such a process could contribute 
to the large amount of hydrogen detected in the WR star,  although one would need to account for the secondary (which is observed to be an O-type supergiant) 
not entering the WR phase as a result of mass-loss during RLOF.

All binary solutions found go through a RLOF phase before the primary reaches the WR phase, which could already be anticipated given 
the large radii reached by the primaries after leaving the main sequence (cf.\, Table\,\ref{tab:evocompsinnon}).
As  discussed in Appendix \ref{sec:inditracks}, RLOF typically removes $\approx 30\,M_\odot$ from the primary, at times partially accreted by the secondary. Mass transfer thus 
appears to be crucial  for the detailed evolution of the systems which do not experience QCHE.

Despite the importance of mass-transfer, our results  indicate that binary interaction does not contribute to the existing 
number of WR stars in the SMC. In 
Sect.\,\ref{sec:introduction}, we argued that it is a priori expected that the majority (if not all) of the SMC 
WR population would stem from
binary evolution, which generally enables WR stars to form at lower initial masses ($M_\text{i} \gtrsim 20\,M_\odot$) compared with 
single stars ($M_\text{i} \gtrsim 45\,M_\odot$).
And yet, the initial masses of the primaries  are found to be in excess of $60\,M_\odot$.
This means that all WR components had large enough initial masses to become WR stars \emph{regardless} of binary effects. 

It is conceivable that the limit of $M_\text{i} \approx 45\,M_\odot$  for SMC stars to become WR stars is an overestimation, as it strictly holds for non-homogeneous 
stars. This limit can decrease to $\approx 20\,M_\odot$ if homogeneous evolutionary tracks are considered (cf.\ Fig.\,\ref{fig:HRDsin}). Regardless, it is unclear 
why no WR binaries with intermediate-mass (20 - 40$\,M_\odot$) primary progenitors are found. 
Since the initial mass function strongly favors the formation of lower mass stars \citep{Kroupa2001}, 
one would expect to see at least some 
WR binaries originating from intermediate-mass progenitors. 
The only WR stars in the SMC which imply 
intermediate-mass progenitors are the putatively single stars AB\,2 and 10. As showed in Paper I, their HRD positions can be 
reproduced by assuming QCHE. 
Alternatively, they could stem from binary evolution.  While the lack of confirmed companions sheds doubts on this scenario, 
post-RLOF binaries would often appear as single stars due to their small typical velocity amplitudes and/or the large 
brightness contrast of the companions \citep{DeMink2014}. The apparent lack of detected WR stars which are a direct result of
binary interaction could thus be due to an observational bias.

Various studies \cite[e.g.][]{Packet1981, Shara2015} 
suggest that the secondary should be spun up to near-critical rotation velocities 
as a consequence of mass accretion during the primary's RLOF.
Interestingly,  all O-companions are found to have $v_\text{eq}$ values above the average for 
single O stars
($\approx 100\,$\kms, e.g.\ \citealt{Penny1996, Ramirez2013}), yet none of them are near critical ($\approx 500\,$\kms). 
The fact that the O companions rotate with velocities above average
implies that RLOF may have occurred, but the fact that they are sub-critical challenges this scenario. 
A resolution could lie in tidal interactions and/or  mass-loss, which together lead to a rapid loss of angular momentum. Alternatively,
the spin up during RLOF may be overestimated.

\subsection{The strange case of AB\,6}
\label{subsec:WR6}

The system SMC\,AB\,6 stands out as very enigmatic. In Sect.\,\ref{subsec:masses}, we showed that  the
luminosity and orbital mass inferred for the WR component of AB\,6 imply that it greatly exceeds its Eddington limit within errors.
This means one or more of the following: (a) The parameters derived for the system in this 
paper, most importantly $\log L$, are incorrect; (b) The orbital mass derived by FMG is incorrect. 

In Appendix \ref{sec:comments}, we thoroughly describe how the components' luminosities are derived.
One caveat is that the method relies on the adopted abundances. Since we use different elements (C, S, P), and since 
not all are expected to change with evolution, a systematic 
deviation is unlikely. Furthermore, the light ratio cannot be very different than derived here, since 
a reduction of the WR luminosity 
would imply an unrealistic increase of the companion's luminosity \citep[e.g.][]{Martins2005}.
However, a third component could contaminate the spectra.   
quite a few examples exist for false analyses of triple systems which were considered to be binary
\citep[e.g.][]{Moffat1977}. 
The immediate neighborhood of AB\,6 is crowded with 
luminous stars and unresolved sources, which increases the probability for a third component contributing to the 
total light of the system. A third component could lead to a smaller luminosity for the WR component,  although it would be hard 
to account for the $\approx1$\,dex downwards revision in $\log L$ which would be necessary to compensate for the discrepancies.

Another possibility is that the orbital mass is incorrect.
To obtain a mass which fits more reasonably with 
our results, an inclination of $\sim 30^\circ$ would be required. However, assuming that the mass ratio derived by 
FMG is correct, such an inclination would also imply 
a mass of $\sim 170\,M_\odot$ for the O companion, which is unrealistic. Alternatively, it is possible that 
the actual mass ratio is different: The RV curve of the O component in AB\,6 shown by FMG is based on noisy data points 
obtained from the motion of the absorption features in the low resolution optical spectra. A larger RV amplitude for the O star
could lead to a larger orbital mass for the WR component. Moreover, a potential third
source would not only affect the derived luminosities, but could also 
affect the RV measurements of the other two components \citep[see e.g.][]{Moffat1977, Mayer2010}.

Although this short-period binary is a potential candidate for unique behavior patterns, 
its properties, as given here, are impossible to explain within the frame 
of binary evolution. This peculiar system should clearly be subject to further studies.

\section{Summary}
\label{sec:summary}

This study presented a systematic spectroscopic analysis of all five 
confirmed WR multiple systems in the low metallicity environment of the SMC. Together with Paper I, this work provides 
a detailed non-LTE analysis of the complete SMC WR population. 
We derived the full
set of stellar parameters for all components of each system, and 
obtained important constraints on the impact of binarity on the SMC WR population.

Mass-luminosity relations (MLRs) calculated for homogeneous stars \citep{Graefener2011} reveal a good agreement for the very massive components 
of SMC\,AB\,5 (HD 5980).  Because of the errors on $\log L$ and $M_\text{orb}$, it is difficult to tell whether these stars are core H-burning or He-burning, 
although their derived positions are more consistent with core H-burning. The remaining WN stars in our sample show higher luminosities than predicted by the MLR calculated 
for pure He stars, implying core He-burning and shell H-burning (shell He-burning for the WO component in AB\,8). 
This is consistent with the fact that all WN components show traces for hydrogen 
in their atmospheres ($0.1 < X_\text{H} < 0.4$). 

The WO component in AB\,8 is found to slightly exceed its Eddington limit, but this 
is likely a consequence of the errors on $\log L$ and $M_\text{orb}$. The small orbital mass 
($\approx 10\,M_\odot$) and high 
luminosity ($\log L \approx 6.3\,[L_\odot]$) inferred for the WR component of AB\,6 imply that it greatly exceeds its Eddington limit, which is clearly unphysical. 
We believe that the most likely resolution is 
an underestimation of the orbital mass, possibly  because of a third component contaminating the spectrum of the system, which could also affect the 
derived luminosity.
Overall, the positions of the WR stars in our sample on the $M-L$ diagram (Fig.\,\ref{fig:orbmass}), together with the derived atmospheric chemical compositions, 
suggest that the stars are not chemically homogeneous, with the possible exception of AB\,5.

A comparison of the observed properties of each system to evolutionary tracks calculated with the BPASS and BONNSAI tools for chemically 
homogeneous/non-homogeneous single stars suggests that chemically homogeneous evolution (QCHE)  is not consistent with four of the five systems
analyzed (AB\,3, 5, 6, and 7). In the case of AB\,5, this is a direct result of the temperature derived in this study, 
which could be biased 
by the effects of wind-wind collisions (WWC), hindering us from a definite conclusion in its case. There are good reasons to believe that 
the components of AB\,5 did in fact experience QCHE, but not without open problems (see Sect.\,\ref{subsec:evolution} and Appendix 
\ref{sec:inditracks}). The case of AB\,8 is uncertain, as QCHE can explain its evolutionary state, although it is not a necessary assumption. 
This stands in contrast to the putatively 
single SMC WR stars, which are generally better understood if QCHE is assumed. The difference presumably stems from tidal synchronization, inhibiting 
an efficient chemical mixing in the stars.  We showed that, if QCHE is avoided, the components of all our analyzed systems 
had to have interacted via mass-transfer in the past.

Mass-transfer in binaries is found to  strongly influence the detailed evolution of the SMC WR binaries, 
significantly changing and redistributing the total mass of the system. That said, stellar winds too play a significant role in 
determining the final masses of the components, which stresses the importance of accurate mass-loss calibrations in 
evolutionary codes.

Despite the importance of mass-transfer, initial masses derived for the primaries are in excess of $60\,M_\odot$,
well above the lower limit for single stars to enter the WR phase at SMC metallicity.
Put differently, it seems that the primaries would have entered the WR phase regardless of binary effects. 
This suggests that the existing number of WR stars in the SMC is not increased  
because of mass transfer, in agreement with the fact that the observed WR binary fraction in the SMC 
is 40\,\%, comparable with the MW.  No WR binaries are found with intermediate-mass ($20-40\,M_\odot$) progenitors, 
although their existence is predicted by stellar evolution models. Since post-RLOF systems tend to appear 
as single stars, this could  
be due to an observational bias.

The sample clearly suffers from low number statistics, and so any general claims put forth in this paper should
be taken with caution. Our understanding of binary effects on the evolution of massive stars 
is expected to improve in the near future, as the sample of analyzed WR binaries 
will continue to grow. 
Meanwhile, our results should serve as a beacon for stellar evolution models aiming at reproducing the observed 
statistical properties of the plethora of massive stellar objects, to which WR stars belong.

\begin{acknowledgements}
We thank our anonymous referee for their constructive comments.
TS is grateful for financial support from the Leibniz Graduate School for Quantitative 
Spectroscopy in Astrophysics, a joint project of the Leibniz Institute for Astrophysics Potsdam (AIP) 
and the institute of Physics and Astronomy of the University of Potsdam. 
LMO acknowledges support from DLR grant 50 OR 1302. AS is supported by the Deutsche Forschungsgemeinschaft under grant HA 1455/26.
AFJM is grateful for financial support from NSERC (Canada) and FRQNT (Qu\'ebec).
JJE thanks the University of Auckland for supporting his research. JJE also wishes to acknowledge the contribution of the NeSI
high-performance computing facilities and the staff at the Centre for eResearch at the University of Auckland.
We thank F. Tramper for providing us 
a reduced spectrum of AB\,8. TS acknowledges helpful discussions with G.\ Gr\"afener and N.\ Langer. 
This research made use of the SIMBAD and VizieR databases, operated
at CDS, Strasbourg, France.  
\end{acknowledgements}

\bibliography{literature}

\Online

\begin{appendix}

\section{Comments on individual targets}
\label{sec:comments}

In the few paragraphs below, we give a short overview on each system, and 
discuss specific issues related to their analysis. We also include an overview of the X-ray
properties of each system.

\emph{\bf AB\,3:}
This system is classified as WN3h + O9 and is found to have a period of $P = 10.0\,$d \citep[][, FMG]{Moffatb}.
An X-ray source (\object{CXOU J004959.4-732211}) was detected in the close 
vicinity of AB\,3 by both {\em Chandra} and {\em XMM-Newton} at a separation 
of $2.9''$. Within position uncertainties, the source could coincide with AB\,3.
If so, the X-ray luminosity of AB\,3 is 
$L_{\rm X}\approx 2\times 10^{33}$\,erg\,$\text{s}^{-1}$.

The spectrum is clearly dominated by the WR component; He\,{\sc i} lines belonging to the secondary are only barely visible. 
The light ratio could be roughly constrained using the temperature-insensitive Balmer lines, and the 
implied luminosity obtained for the O companion suggests it is an O9 dwarf. The temperature of the WR star could be fairly well constrained due 
to the clear signatures of the C\,{\sc iv} resonance doublet in the UV and the strong N\,{\sc v} features in the optical, as well as the weak 
N\,{\sc iv} features. $R_\text{t}$, $v_\infty$ and $D$ follow by fitting the strengths 
and shapes of the emission lines. 
We find a significant hydrogen content in the primary's atmosphere. 

\begin{figure}[!htb]
\centering
  \includegraphics[width=\columnwidth]{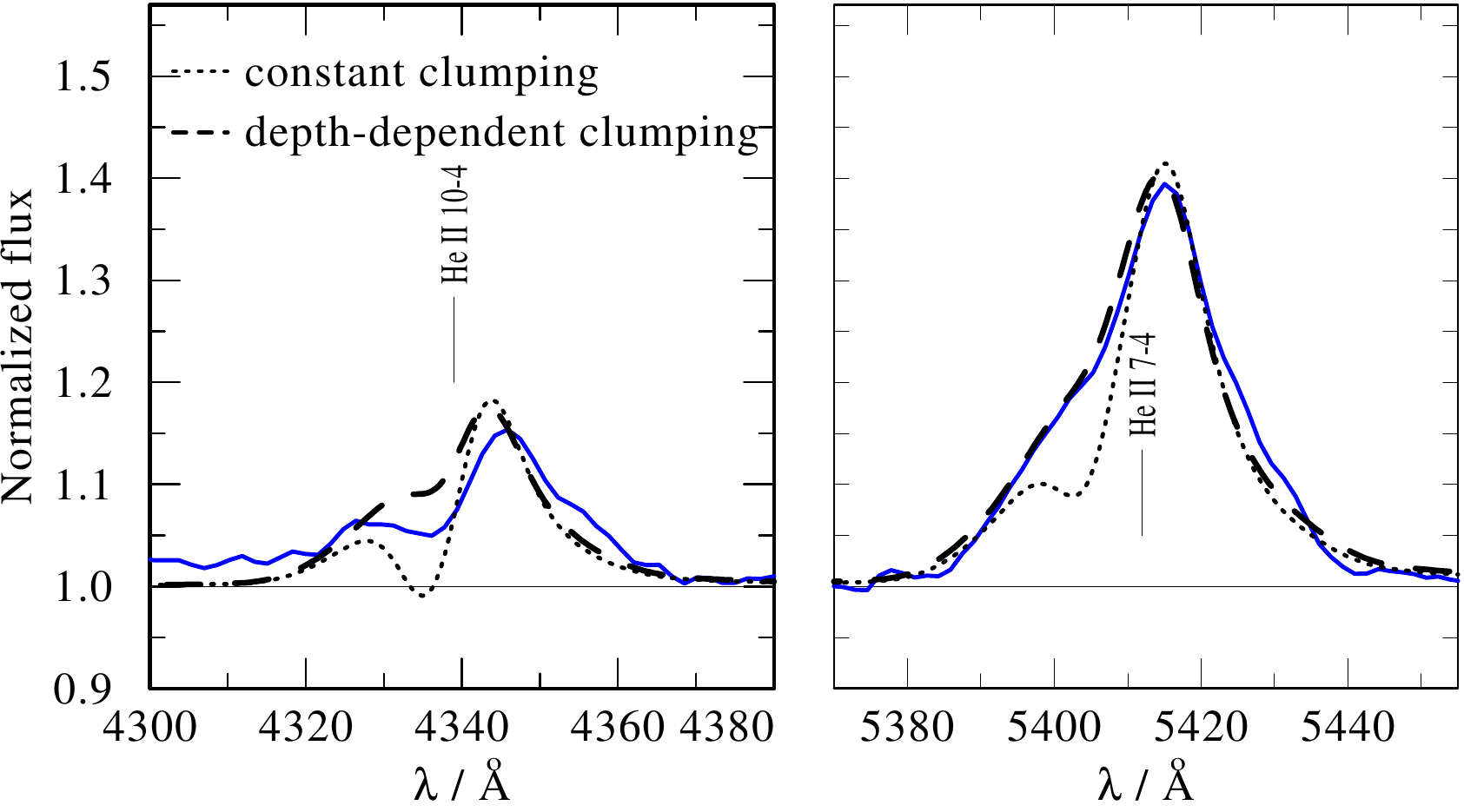}
  \caption{Two models corresponding to the WR model of AB\,3 (cf.\ Table\,\ref{tab:stellarpar}) with a constant      
  clumping factor $D(r) = D = 10$ (black dotted line) and a depth-dependent clumping factor, where $D(R_*) = 1$ and $D(r) = 10$ 
    at $v(r) \sim 0.5\,v_\infty$ (black dashed line) compared to observations (blue solid line). 
    We note that the O model contributes very little to these lines and is therefore omitted from this figure for clarity.
       }
\label{fig:DDclump}
\end{figure}

The emission visible at $\lambda \sim 4640$ probably corresponds to the N\,{\sc iii} doublet 
$\lambda \lambda\,4634, 4640$. No WR models are capable of reproducing this emission simultaneously with the N\,{\sc iv} 
and N\,{\sc v} features. It could only be reproduced by assuming higher temperatures and lower gravities for the O companion, and 
an extremely high N abundance, of the order of 30 times the SMC value. 
The potential nitrogen enrichment is given as an alternative $X_\text{N}$ value in Table\,\ref{tab:stellarpar}.
Alternatively, this feature could arise from WWC \citep[e.g.][]{Rauw1999}.
The companion's terminal velocity is roughly estimated based on the shape of the badly resolved C\,{\sc iv} 
resonance doublet, while 
the mass-loss rate is adopted from \citet{Vink2000}.

\emph{\bf AB\,5:} 
This strongly variable quadruple system, also known as HD 5980, contains four companions: an eclipsing binary system 
composing two WR stars (A:WNN6h + B:WN6-7) 
with a period of $P = 19.3$\,d, 
and another, probably distant binary system (C: O7~I + ?), 
responsible for distinct and relatively static absorption features in the spectrum \citep{Breysacher1982, Niemela1988,
Koenigsberger2010, Koenigsberger2014}. 
Here, we treat star C as a single star, since the fourth component is not noticeable in the spectrum, and  
therefore should not influence our results.
The primary star went through a powerful LBV-type eruption in the year 1994. Since the eruption, the primary gradually increased in 
temperature from $\sim 25$ to $\sim 50$\,kK, while
its mass-loss rate has been gradually decreasing from $\sim 10^{-3.7}$ to $\sim 10^{-4.4}\,M_\odot\,{\rm yr}^{-1}$, 
and the terminal velocity increasing from $\sim 500$ to $\sim 2400$\,\kms \citep{Georgiev2011}. 
The system is the most X-ray luminous WR star in the SMC, with an X-ray luminosity $L_{\rm 
X}\approx 10^{34}$\,erg\,${\rm s}^{-1}$. Orbital variations in its X-ray flux 
were reported and could be explained by WWCs between the 
components A and B \citep{naze2007}.   AB\,5 is a prominent candidate for producing  
a massive black hole binary system such as recently detected by LIGO via gravitational waves \citep{Abbott2016, Marchant2016}.

\citet{Georgiev2011} presented a thorough analysis of the stars A and C at different epochs since the 1994 eruption, taking 
advantage of the fact that, at least for optically thick lines, the eclipse at phase $\phi = 0$ (star A in front of star B) can be 
assumed to be full \citep{Perrier2009}. In our analysis, we use only the most recent spectra available. 
While the spectra taken during primary eclipse 
were taken at different epochs between the years 2002 and 2009 (see caption of Fig.\,\ref{fig:AB5bin}), 
the system was stable during this period \citep{Georgiev2011}, so fitting them simultaneously is justified. 
Using the same approach as \citet{Georgiev2011} to isolate the stars A and C during the eclipse of B ($\phi = 0$), we find similar parameters 
for both stars, although our solution implies a lower temperature for the primary ($T_* =$45\,kK vs.\ 60\,kK). 
Stars of subtype WN6h can have temperatures 
ranging between these values \citep[e.g.][]{Hamann2006}.

We note that 
we neglect the effects of WWC here. \citet{Moffat1998} claim that the phase-dependent width of the strongest optical emission lines is dominated by 
WWC during the eruptive era. Since the eruption, the primary's mass-loss has significantly decreased, so WWC are now not expected to be as dominant, but could 
still contribute to the spectrum. Indeed, we find the strong He\,{\sc ii} $\lambda 4686$ line is broader at maximal width (quadrature) than can be reproduced by 
a superposition of two WR stars with the given velocity amplitudes. This does not hold for the weaker He\,{\sc ii} and nitrogen lines, partly 
because they form closer to the stellar surface, as \citet{Koenigsberger2014} noted. While He\,{\sc ii} $\lambda 4686$
could be contaminated by WWC, our results do not depend on this line. However, WWC could potentially influence the strong He\,{\sc i}\,$\lambda 5876$
line, which was used to constrain $T_*$ and $\dot{M}$. The discrepancy in $T_*$ could therefore be partly due to WWC.

We can only perform a rough analysis of the out-of-eclipse spectra, since the most recent publicly available observations 
covering the range $1200-2000\,\AA$ were taken in 1999, closer to the eruption. We thus rely on previous 
analyses as much as possible.
\cite{Perrier2009} performed a quantitative analysis of the system's light curve taken in 1979, before the 1994 eruption, 
and derived helpful constraints. They found that the relative contributions of stars A, B, and C in 1979 are approx.\ 
0.4, 0.3, and 0.3 in the visual, respectively. \cite{Foellmi2008} present a comparison between this light curve and newer light curves 
taken in the years 2005 and 2006. While \cite{Foellmi2008} noted some interesting differences between the pre- and post-eruption light curves, 
the global behavior and strengths of flux minima during eclipse are virtually unchanged. This implies that the results published by 
\cite{Perrier2009} hold now as well. Furthermore, the fact that both primary and secondary minima are of very similar strengths implies that 
stars A and B have very similar temperatures. The clear appearance of N\,{\sc iv} and N\,{\sc v} lines which can be attributed to both components 
in the out-of-eclipse spectra confirms this. With the parameters of stars A and C fixed and the temperature of star B fixed to that of star A, 
the visual light ratios derived by \cite{Perrier2009} imply the luminosity of star B, and thus its radius.
Finally, with these parameters fixed, we analyze the available out-of-eclipse spectra to derive the secondary's wind parameters. 
Applying a 3D integration technique \citep{Shenar2015}, an upper limit on the components' 
$v \sin i$ can be constrained using the N\,{\sc iv}\,$\lambda 4058$ line, which forms very close to the stellar surface and is thus sensitive 
to surface rotation.

\emph{\bf AB\,6:} 
With a period of $P = 6.5\,$d, this SB2 WR + O binary system, classified WN4 + O6.5 I by FMG, 
is the shortest period WR binary known in the SMC 
\citep{Moffat1982, Hutchings1984}. Despite its short period, 
the system shows no significant photometric variability.  
AB\,6 was detected in the {\em XMM-Newton} survey \citep{Laycock2010, Sturm2013} with an
X-ray luminosity of $L_{\rm X}\approx 7\times 10^{33}$\,erg\,$\text{s}^{-1}$ (in the 0.2-4.5\,keV band). The source was found to be 
variable, with the $L_\text{X}$ reaching  a peak of $\approx 10^{34}$\,erg\,s$^{-1}$.

The luminosity of the WR component turns out to be very high, 
a fact which is not easy to explain given 
its orbital mass (see Sect.\,\ref{subsec:masses}). Since this is a direct result of the 
derived light ratio, we briefly discuss its determination. 
The light ratio was derived on the basis of 
mainly three features: The P\,{\sc v}\,$\lambda\,1128$ and S\,{\sc iv}\,$\lambda 1073$ resonance lines, 
as well as the 
C\,{\sc iii} $\lambda 1176$ multiplet, all of which are observed in the FUSE domain (see Fig.\,\ref{fig:CIIIhelp}).

\begin{figure}[!htb]
\centering
  \includegraphics[width=\columnwidth]{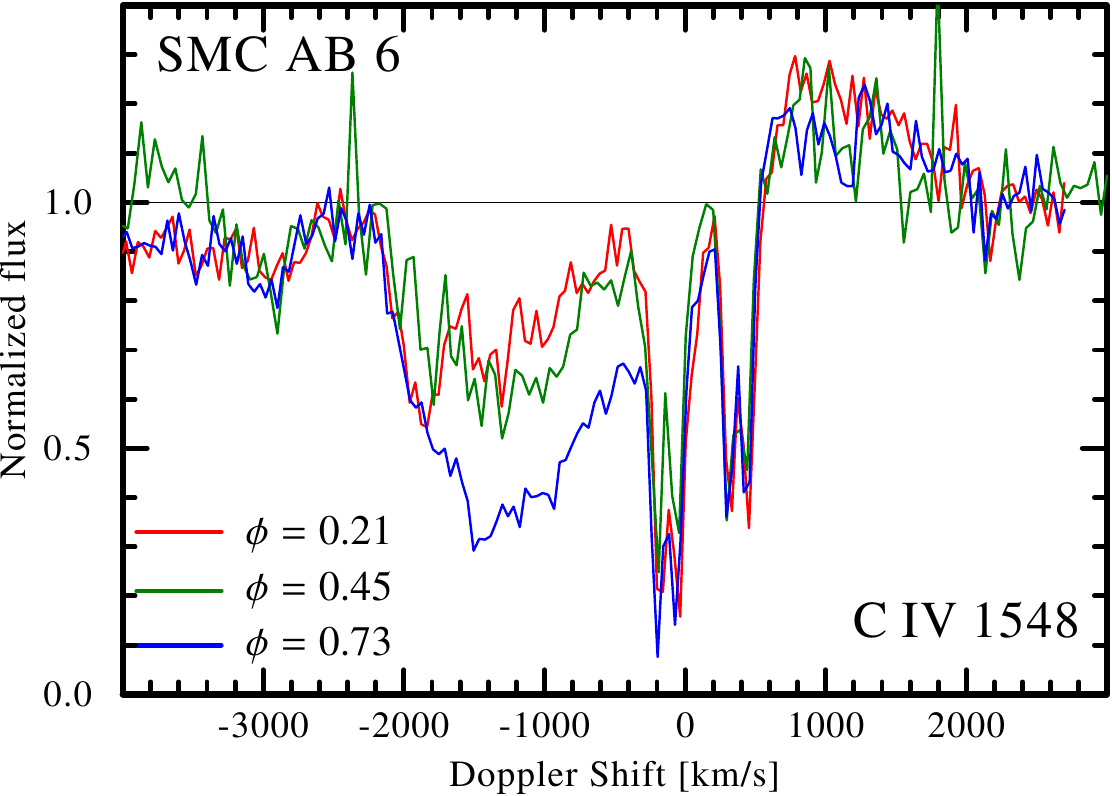}
  \caption{Two HST spectra taken in 1992 (red line: Z0Z30402T, $\phi = 0.21$, blue line: Z0Z30602T, $\phi = 0.73$) 
  and an IUE spectrum taken in 1991 (green line, sp41784, $\phi = 0.45$) are plotted, 
   showing the C\,{\sc iv} resonance doublet
  of AB\,6 in velocity space relative to the blue component at $\lambda 1548$.}
\label{fig:CIVWR6var}
\end{figure}

Moreover, while we do not identify a "step" in the C\,{\sc iv} resonance line (as in 
AB\,5, see Fig.\,\ref{fig:CIVhelp}), 
it is clearly unsaturated at all phases, suggesting that it originates in only one of the binary components. We argue that this line 
originates in the O star.
In Fig.\,\ref{fig:CIVWR6var},
we show two HST
 spectra at $\phi = 0.21$ (red line) and $\phi = 0.77$ (blue line), as well as one IUE spectrum
at $\phi = 0.45$ (green line) in velocity space relative to the blue component of the C\,{\sc iv} resonance 
doublet ($\lambda 1548$). The spectra are shifted by $-200\,$\kms\, to account for the  systemic velocity of the system 
(FMG). While the line shows significant variability whose origin should be subjected to further study,  
no obvious radial velocity variation with phase can 
be seen in the line. If the line originates in the WR component, 
there should be a difference of roughly $500$\,\kms\, between 
the velocities of the lines at phases $\phi = 0.21$ and $\phi = 0.73$, but no shift is observed. While emission 
lines need not strictly follow the orbital motion, some shift would be expected at such an RV amplitude. We conclude 
that this line originates in the O star, whose RV amplitude is $\sim 4$ times smaller, 
and that the WR star does not show this line at all, 
as is typical for early WN stars.
This line helps us determine the light ratio as well, and puts constrains on the wind parameters of the O star.
Lastly, the light ratio is consistent with He\,{\sc i} and Balmer absorption lines in the optical spectrum.

As in AB\,3, we find a clear signature of N\,{\sc iii} $\lambda 4640$, which is very unlikely to originate in the WR component given its 
spectral class and derived temperature; indeed, we could not produce N\,{\sc v} at its observed strength simultaneously to any N\,{\sc iii}
features. To reproduce it, an N-enrichment of roughly ten 
times the typical SMC value is assumed in the model of the O component, noted as an alternative value in Table\,\ref{tab:stellarpar}. This feature could also 
originate in WWC, however.

\emph{\bf AB\,7:}
Classified as WN4 + O6 I(f), this SB2 binary has a detected period of $P = 19.6$\,d (FMG).
The system was also observed with {\em XMM-Newton} and is found to have a smaller X-ray luminosity than 
that of AB\,6 ($L_{\rm X}\approx 4\times 10^{33}$\,erg\,$\text{s}^{-1}$), which is also found to be 
variable \citep{Guerrero2008}. Interestingly, although the mass-loss rates of WR stars in the SMC 
are on average lower than in those of their Galactic counterparts, 
the colliding wind SMC binaries (AB\,5, 6, and 7) do not appear to be weaker X-ray sources than 
the Galactic ones. This suggests that the components' mass-loss rates and the resulting
X-ray luminosities may be related in a non-trivial way, or may even be uncorrelated \cite[e.g.][]{Ignace2000}.

The temperature
of the WR component is hard to constrain;
The N\,{\sc v} resonance line 
at $\lambda 1240$ is visible, but could in principle arise in both companions. There is a clear signature 
of N\,{\sc iii}, but \cite{Niemela2002} 
associate it with the O companion based on RV shifts, and there are no N\,{\sc iv} lines observed. Very weak N\,{\sc v} lines in the 
optical can be seen, and suggest that the WR companion is indeed hot and suffers from strong line dilution, similar
to AB\,6. To avoid strong N\,{\sc v} lines in the optical (which are not observed), 
one could either reduce the N abundance to abnormally 
low values (roughly a factor $0.3$ times the abundance given in Table\,\ref{tab:stellarpar}), or increase 
the temperature to $\sim 10^5$\,K, 
where the N\,{\sc v} lines become very weak (because of dominance of the N\,{\sc vi} ion). The latter solution 
provides a better fit to the available data.

Most peculiarly, the C\,{\sc iv} resonance
line only shows a narrow absorption, likely blended with interstellar absorption. 
For this reason, this line cannot help us to deduce the light ratio. As in the case of AB\,6, we use 
the P\,{\sc v}\,$\lambda\,1128$ and S\,{\sc iv}\,$\lambda 1073$ resonance lines, as well as the 
C\,{\sc iii} $\lambda 1176$ multiplet to constrain the light ratio, and derive the remaining parameters for 
both components based on the strengths of He\,{\sc i} and He\,{\sc ii} lines in the optical.
To reproduce the unusual C\,{\sc iv} feature (or, more accurately, lack thereof), we assume a low mass-loss rate for the O-star companion, 
and include X-rays in its model such that the total X-ray luminosity observed in this system is reproduced. 
The X-rays strongly depopulate the C\,{\sc iv} ground state via Auger ionization.  

As in the case of AB\,3 and 6, the N\,{\sc iii} emission excess at $\lambda 4640$ can be reproduced if 
a nitrogen abundance of roughly a factor 10 times the typical SMC abundance is assumed in the secondary. 
In fact, \cite{Niemela2002} showed that the feature roughly follows the orbital velocity of the O companion.
We thus find possible signatures for nitrogen enrichment in SMC\,AB\,3, 6, and 7. If true, this can be a consequence 
of either contamination by the primaries' nitrogen-rich winds, or accretion during the primaries' RLOF. Alternatively, this 
feature could originate in colliding winds, which are likely present in all of these systems.

\emph{\bf AB\,8:}
This system contains the only non-WN star known among the SMC WR sample. It was reclassified several times, 
most recently as WO4 + O4 V by \citet{Bartzakos2001}. \cite{Moffat1985} inferred an orbital period 
of $P = 16.6$\,d for the system. Interestingly, AB\,8 is the only WR binary in our sample not detected in X-rays. 
The star was in the field of 
view of {\em Chandra} and {\em XMM-Newton} observations \citep{Oskinova2013}, 
but only an upper limit of $ 5 
\times 10^{32}$\,erg\,s$^{-1}$ could be established for its X-ray luminosity. Given the very strong wind of the primary, it is 
difficult to explain such a low X-ray luminosity; \citet{Zhekov2015} 
recently reported that the X-ray luminosity of the Galactic WO binary \object{WR 30a} 
is $>10^{34}$\,erg\,s$^{-1}$. In fact, \cite{St-louis2005} report clear signatures 
of colliding winds in AB\,8. While there are possible resolutions to this problem
(e.g.\ large opacity of the surrounding material), 
it remains difficult to explain why AB\,8 does not exhibit a high X-ray luminosity. 

The light ratio of the components 
could be fairly well deduced by taking advantage of the fact that the N\,{\sc v} resonance line clearly 
belongs to the O companion (see Fig.\,\ref{fig:CIVhelp}). The light ratio is consistent with the C\,{\sc iv} resonance line and with the 
other clear features belonging to the secondary, e.g. Balmer and He lines in the optical. The terminal velocity of the companion is found to be comparable to that 
of the primary. The ratio of carbon to helium in the WO component is constrained 
using the diagnostic line pair He\,{\sc ii}\,$\lambda 5412$; C\,{\sc iv}\,$\lambda 5471$, whose ratio 
is primarily sensitive to $X_\text{C} / X_\text{He}$ \citep{Koesterke1995}. The abundances derived 
are generally supported by the remaining carbon and oxygen features.

There is a very strong $\dot{M}$-$T_*$ degeneracy for the WR companion which arises from the fact that, for optically thick winds, 
one could arbitrarily vary the temperature and the mass-loss rate simultaneously to obtain a "photosphere" ($\tau \sim 2/3$) which 
has approximately the same effective temperature and emitting surface \citep[see Fig.\,4 in][]{Todt2015}.
The temperature is clearly higher than $\sim 120$\,kK, because lower values cannot 
reproduce the optical O\,{\sc vi} $\lambda \lambda\,3812, 3835$ emission. For $T_* \gtrsim 120\,$kK,
the degeneracy just mentioned makes it possible to reproduce the O\,{\sc iv} $\lambda 3400$ line regardless of the value 
of $T_*$, as long as  $\dot{M}$ is large enough. Temperatures in the vicinity of 200\,kK produce
emission of the O\,{\sc vi} $\lambda \lambda\,1032, 1038$ resonance doublet that is too strong. $T_* \approx 150$\,kK is chosen as a compromise. 
Note that despite the very large uncertainty in $T_*$, we can constrain $T_{2/3}$ fairly well, since all well-fitting models reach a similar 
temperature at their ``photospheres''.

From the profiles of the O\,{\sc iv}, O\,{\sc v} and O\,{\sc vi} lines in 
the optical, it is clear that the O\,{\sc iv} line is formed far out (flat topped profile), while the O\,{\sc v} and O\,{\sc vi} features 
form closer in. At any temperature within the range mentioned above, 
reproducing the O\,{\sc iv} emission requires large $\dot{M}$ values which produce too strong emission for most features in the spectrum 
(especially the strong C\,{\sc iv} lines in the UV and optical and the He\,{\sc ii} $\lambda 4686$ line),
while strongly weakening the O\,{\sc vi} emission in the optical below the observed value. 
\cite{Tramper2013}, who were also unable to simultaneously fit both O\,{\sc iv} and O\,{\sc vi} lines at their observed strengths, suggest that
this discrepancy may be related to soft X-ray K-shell ionization. However, 
we are unable to verify this claim. \cite{Zsargo2008} show that O\,{\sc vi} 
could be sensitive to a hot, inter-clump medium, not accounted for in our models currently.  Overall, 
the multitude of lines belonging to carbon and oxygen make the problem of modeling WO stars more difficult, and the fit quality is indeed 
inferior to that of other systems (cf.\ Appendix\,\ref{sec:specfits}). Our fit represents a compromise to the most important features which 
could be recognized in the spectrum of AB\,8.

\cite{St-louis2005} studied variability in the system and obtained helpful constraints on the stellar parameters of both companions. 
The mass-loss rates derived for both components are in agreement (within errors), as are the radii of both stars. 
One very strong discrepancy lies in the bolometric luminosity ratio of both companions: \cite{St-louis2005} infer $L_\text{O} / L_\text{WR} = 3.5$ by 
modeling the variability of the O\,{\sc vi} resonance line, while 
we find $L_\text{O} / L_\text{WR} \approx 0.5$ from our spectral modeling. 
Assuming the luminosity and radii ratio inferred by \cite{St-louis2005}, however,
the Stefan-Boltzmann relation implies $T_\text{WR} \sim 1.35 T_\text{O}$.  For a typical temperature
of an O4 V star ($45$\,kK; \citealt{Martins2005}), 
one would find a temperature of only $T_\text{WR} \approx 60$\,kK for the WO component, which is less than 
half the derived temperature, and much lower than temperatures derived 
for other WO stars \citep[e.g.][]{Tramper2013}. We therefore believe that our results are more consistent.

Another interesting spectral feature which we could not reproduce is the strong absorption seen in the FUSE spectra 
around $\lambda \approx 1160\,\AA$. This absorption feature is persistent in all observations and follows the orbital 
motion of the WO component. It likely belongs to either to C\,{\sc iv} or C\,{\sc iii}, but its strength is much larger than 
what can be reproduced by the models, especially given the deduced light ratio. It is possible that this feature is a manifestation of 
the components' interaction.

\section{Following the binary evolution of each system}
\label{sec:inditracks}

In this section, we discuss the evolutionary path of each system in more detail,  as inferred from the corresponding binary evolution tracks.
We note that the
best-fitting tracks should be at best seen as an approximation to the system's evolution, as we work with a coarse grid 
of evolutionary models. Calculating individual tracks for the analyzed systems is beyond the scope of this paper.

{\bf AB\,3:}  The solution to this system is found to reproduce all observable quantities, and is not sensitive to 
small changes in the weighting (i.e.\ changes in $\sigma_n$).
The system starts off with $M_1 = 60\,M_\odot$ and $M_2 = 20\,M_\odot$,   
an initial period of $\approx 40\,$d, and a separation of $200\,R_\odot$. The primary
is thus already born very massive and luminous ($\log L_1 \sim 5.7$) while being on the main sequence, 
whereas the secondary's parameters correspond to an O\,7 dwarf. After about $3.8$\,Myr, 
the primary becomes a supergiant and reaches a radius of roughly  $100\,R_\odot$, at which point 
it fills its Roche lobe and a case B (shell H-burning) mass-transfer phase initiates, during which the primary
rapidly loses mass in a non-conservative fashion  because of the secondary's 
inability to accrete matter \cite[see Sect.\ 2.2.2 in][]{Eldridge2008}.
During the RLOF phase, which 
lasts $\sim 4\cdot10^4$\,yr, the mass of the primary sharply drops from $55$ to $25\,M_\odot$, 
while the secondary maintains a roughly identical mass of $\sim20\,M_\odot$. 
During the mass-transfer phase, the period drastically decreases from its initial value to $\sim10$\,d, while the separation decreases 
to $\sim 55\,R_\odot$. It lasts roughly $10^5$\,yr until the system evolves to its current state, roughly $3.9$\,Myr after its birth.

{\bf AB\,5:}  The best-fitting solution reproduces all 
parameters within a 2\,$\sigma$ level, although the components' masses are overpredicted by $\approx 1.5\sigma$. 
According to the best-fitting track, the system starts off with very large masses of $M_1\ = 150$ and $M_2 = 75\,M_\odot$, an initial period 
of $\approx16\,$d and a separation of about $160\,R_\odot$. 
Strong stellar winds remove 
more mass from the system during its evolution.
After roughly 2.3\,Myr, the primary reaches a radius 
of $\approx80\,R_\odot$ and, with the primary still core hydrogen burning, 
a partially conservative case A RLOF phase initiates for a short duration of $\approx2.5\cdot10^4$\,yr, during which 
roughly $25\,M_\odot$ are transferred from the primary onto the secondary. The period and separation do not change significantly during this time.
We observe the system ``shortly'' hereafter at an estimated 
age of $2.6\,$My, with the primary's mass now $\approx 85\,M_\odot$ and the secondary's $\approx 95\,M_\odot$.

By assigning more weight to the masses  during the fitting procedure (e.g.\ by making $\sigma_\text{M}$ artificially smaller), a different solution is found 
with $M_\text{i} = 120\,M_\odot, q_\text{i} = 0.3$, and $P_\text{i} = 10$\,d, which reproduces the masses better (difference < 1$\sigma$), 
but underpredicts the period by more than $2 \sigma$ and the secondary's luminosity by $1.5\sigma$.
Generally, the solution space for this system is very non-linear and sensitive to small 
variations of $\sigma_n$, with the initial mass of the primary ranging between $100\,M_\odot$ and $150\,M_\odot$.
Given that this system is currently in 
a highly variable, rapidly changing evolutionary state,
it is perhaps not surprising that we find it difficult to constrain a unique solution in this case. 

\cite{Koenigsberger2014} claimed that RLOF is unlikely to have occurred in this system due to the very similar 
masses and hydrogen content of both companions. These authors developed an evolutionary model that avoids 
a RLOF phase by assuming QCHE,  and obtain results similar to those given in this study for the homogeneous 
case (Table \ref{tab:evocompsinhomo}). In this scenario,
both components 
move along a diagonal towards the upper-left corner of the HRD after leaving the main sequence, 
remaining compact throughout their evolution. In this study, the QCHE scenario does not fit well to the system (cf.\ Tables\,\ref{tab:evocompsinnon} and 
\ref{tab:evocompsinhomo}). The reason 
is that \citet{Koenigsberger2014} adopt a temperature which is higher than derived here (45\,kK vs. 60\,kK). A higher 
temperature is more consistent with QCHE (see right panel of Fig.\,\ref{fig:HRDsin}). As we discuss in Appendix\,\ref{sec:comments}, 
WWC in the system lead to uncertainty in the true temperature of the primary and secondary.

 We find several binary solutions which successfully produce similar post-RLOF masses at roughly the observed 
$T_*$ and $\log L$ inferred for the primary and secondary. However, as discussed in Sect.\,\ref{subsec:evolution}, although 
the BPASS code does not deliver the scondary's hydrogen abundance, it would require some fine-tuning to obtain scenarios which include mass-transfer and which would 
predict a significant hydrogen depletion for both components in the system simultaneously.  
All things considered, QCHE without mass-transfer seems 
to offer a more natural explanation to the current state of the system.

However, 
it is noteworthy that the primary has in fact been observed at much lower temperatures of about $20\,$kK 
during the 1994 eruption \citep{Georgiev2011}, while the QCHE scenario suggested by \cite{Koenigsberger2014} implies that the 
primary's temperature 
continuously increased from its initial value of $\approx 50\,$kK. \cite{Koenigsberger2014} suggest that the 1994 erruption may involve 
an instability arising from an increase of the full Eddington Gamma $\Gamma$ beyond unity. Indeed, such instabilities are not dealt with 
in the current generations of codes. And yet, since we know that the star \emph{did} travel towards the right side of the HRD, 
it is dubious to assume that no mass-transfer occurred throughout the system's evolution. In fact, the solution offered by \cite{Koenigsberger2014} suggests 
that a similar instability should have occurred in star B
even earlier for the same reason.  Moreover, one would have to explain the apparent discrepancy between 
the necessary near-critical ($\sim 400- 500\,$\kms) rotation rates needed to induce QCHE and the velocities implied by orbital synchronization ($50-100\,$\kms).
To obtain a definite answer to this question, accurate modeling of this system, which would account for rotation, tidal forces, and mass-transfer from both components 
consistently, is necessary. This is beyond the scope of the current paper.

{\bf AB\,6:} As could be anticipated, no ordinary binary evolutionary track can reproduce the peculiar combination of 
large luminosity ($\log L \sim 6.3$) and hydrogen content ($X_\text{H} \sim 0.4$) simultaneously with the small orbital mass ($9\,M_\odot$) of 
the primary component.  For this system, we therefore ignored the primary's orbital mass in the fitting procedure.

The system was originally composed of two massive stars with $M_1 = 100$ and $M_2 = 50\,M_\odot$ at a short 
initial period of $P \approx 6\,$d and an initial separation of $\approx 80,R_\odot$. As the primary starts to fill its Roche lobe, 
roughly 2.5\,Myr after its formation, a case A RLOF phase initiates, lasting for a long period of
$\sim 4\cdot10^5$\,yr. During this time, $\approx 20\,M_\odot$ are removed from the primary to the secondary, and the primary's hydrogen
content drops from $X_\text{H} = 0.7$ to $X_\text{H} = 0.35$. Strong stellar winds are responsible for a further decrease of the 
hydrogen content and the stellar masses. The state closest to the observables of this system is  
reached at an age of $\sim 3.0\,$Myr, although the hydrogen content of the primary is underestimated by the track. This is possibly due 
to overestimated stellar winds in the BPASS code. Alternatively, 
RLOF from the secondary onto the primary (not accounted for in the code) could also lead to a hydrogen enrichment.

{\bf AB\,7:} The solution to this system reproduces its observables  well and is not sensitive to small changes in $\sigma_n$.
The system originally comprised of two very massive stars, 
with the primary having twice the mass 
of the secondary: $M_1 = 80$ and $M_2 = 40\,M_\odot$. The initial separation is large, $\approx 250\,R_\odot$, and the initial
period is $\approx 40\,$d. After roughly $3.3\,$Myr, the primary went through a case B RLOF for $\approx 3\,\cdot10^4$\,yr, removing about 
$30\,M_\odot$ from the primary, which is lost from the system because of the secondary's inability to accrete. During RLOF, 
the orbital period is halved, while the separation reduces to $\sim 150\,R_*$.
After $\approx 10^5$\,yr, 
at an age of 3.4\,Myr, the system reaches its currently observed state, with $X_\text{H} = 0.15$ remaining in the outer
envelope of the primary, 
and the secondary becoming an evolved, luminous star.

{\bf AB\,8:} The best-fitting solution reproduces most observables to a satisfactory level. However, the initial mass of the primary is 
sensitive to small changes in the weighting, and ranges between 100 and 150$\,M_\odot$. According to the best-fitting track,
the system was originally composed of two hot and massive stars of masses 
$M_1 = 150$ and $M_2 = 45\,M_\odot$, with a very short initial period of $10\,$d. 
As the primary leaves the main sequence and reaches a radius of $\sim 50\,R_\odot$ after about $2.2$\,Myr, 
it goes through a case B RLOF phase lasting for $\approx 10^5$\,yr, losing $\sim 25\,M_\odot$ in a conservative fashion and transferring it 
onto the secondary, until the masses become $M_1 \approx 100$ and $M_2 \approx 70\,M_\odot$. The period and separation are reduced by $\sim 30\%$ at the end of 
the RLOF phase.
During the next $0.8$\,Myr, the primary suffers 
extreme mass-loss, until it reaches a mass of about 25$\,M_\odot$, while the secondary's mass remained roughly constant. 
This is the current state 
of the system, at an estimated age of roughly 3\,Myr. 

\section{Spectral fits}
\label{sec:specfits}

\begin{figure*}[!htb]
\centering
  \includegraphics[width=\textwidth]{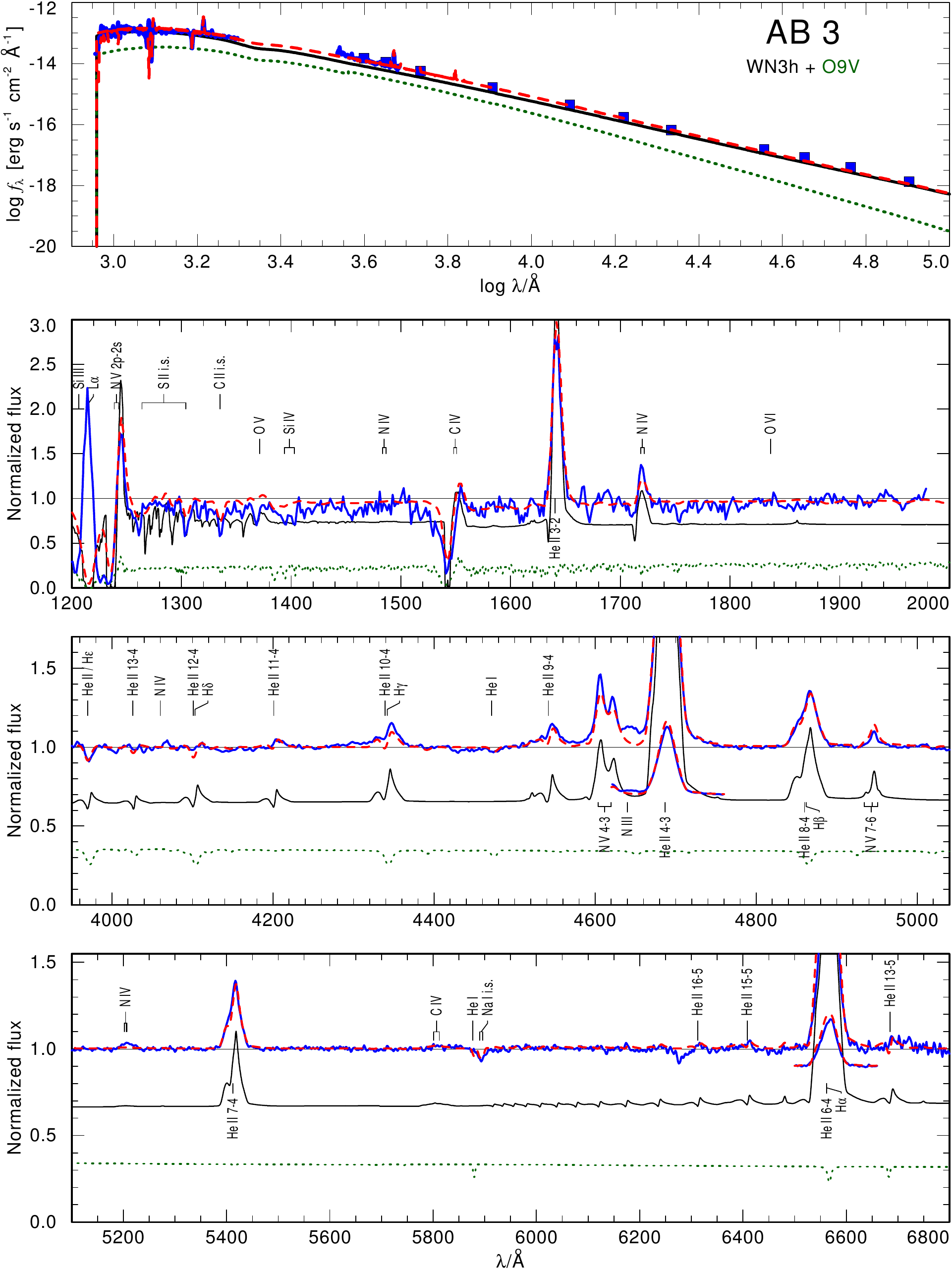}
  \caption{Comparison between the best fitting composite model spectrum (red line) and the observations (blue squares 
    and lines) for the SED (upper panel) and normalized spectra (lower panels) of SMC\,AB\,3. The composite model 
    is the sum of the WR model (black line) and companion model (green line). The observed UV spectrum was taken 
    with IUE in 1988 (ID:sp33457, $\phi = 0.3$), and the optical spectrum are co-added spectra from FMG (see Sect.\,\ref{sec:data}).
    }
\label{fig:AB3}
\end{figure*} 

\begin{figure*}[!htb]
\centering
  \includegraphics[width=\textwidth]{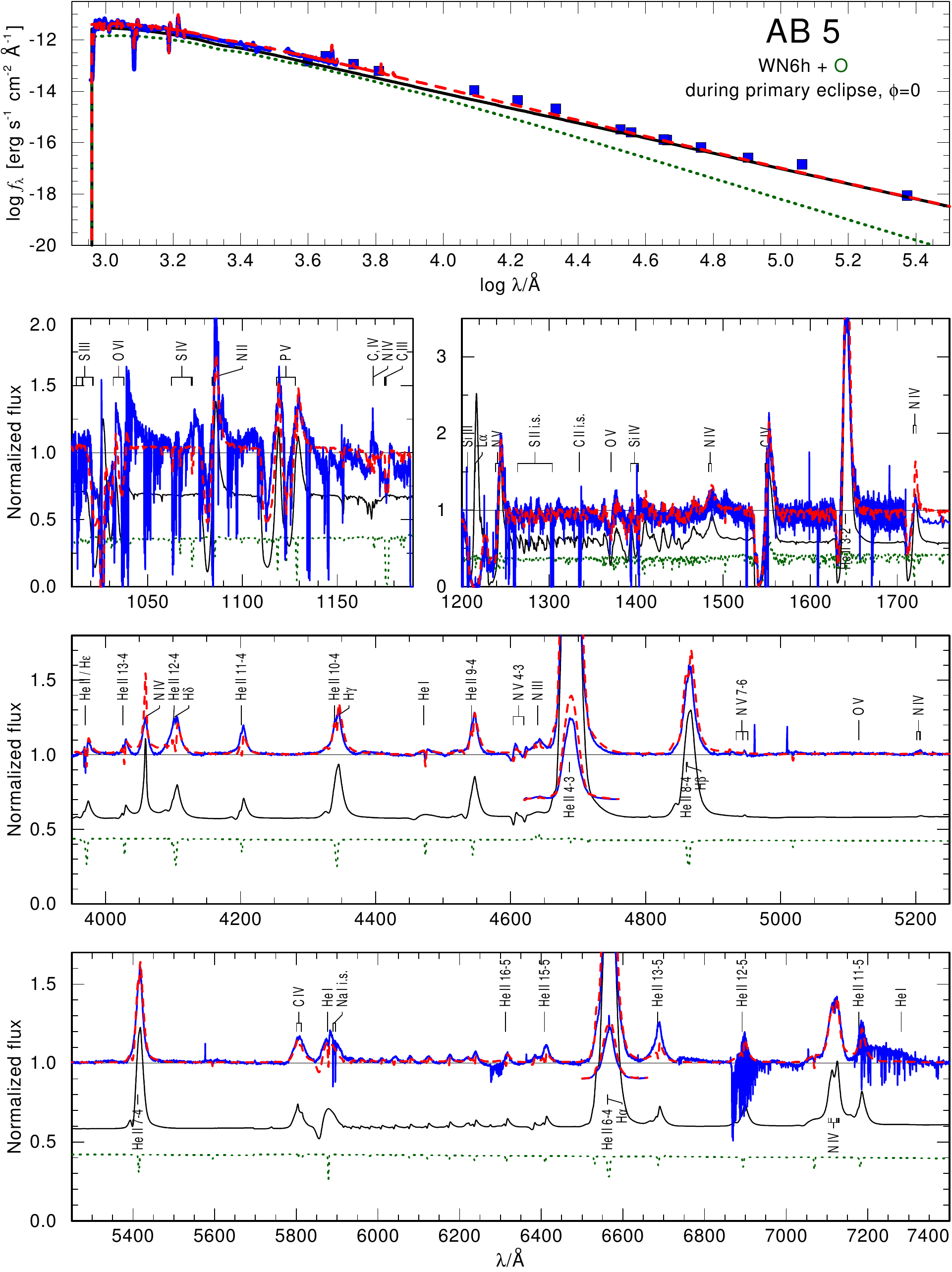}
  \caption{Same as Fig.\,\ref{fig:AB3}, but for SMC\,AB\,5 during an eclipse of the primary by the secondary ($\phi = 0$). The 
  UV spectra shown were taken with FUSE in 2002 (ID:p223010100000, $\phi \eqsim 0$) and HST in 2009 (ID:ob2na1020, $\phi \eqsim 0$). 
  The optical spectrum consists of the co-added FEROS spectra at $\phi \eqsim 0$, described in Sect.\,\ref{sec:data}. 
  Note that most photometry measurements (blue squares) were taken outside 
  eclipse, which is why the total model fails to reproduce their values. Fig.\,\ref{fig:AB5trip} shows a fit to the system 
  outside eclipse.}
\label{fig:AB5bin}
\end{figure*} 

\begin{figure*}[!htb]
\centering
  \includegraphics[width=\textwidth]{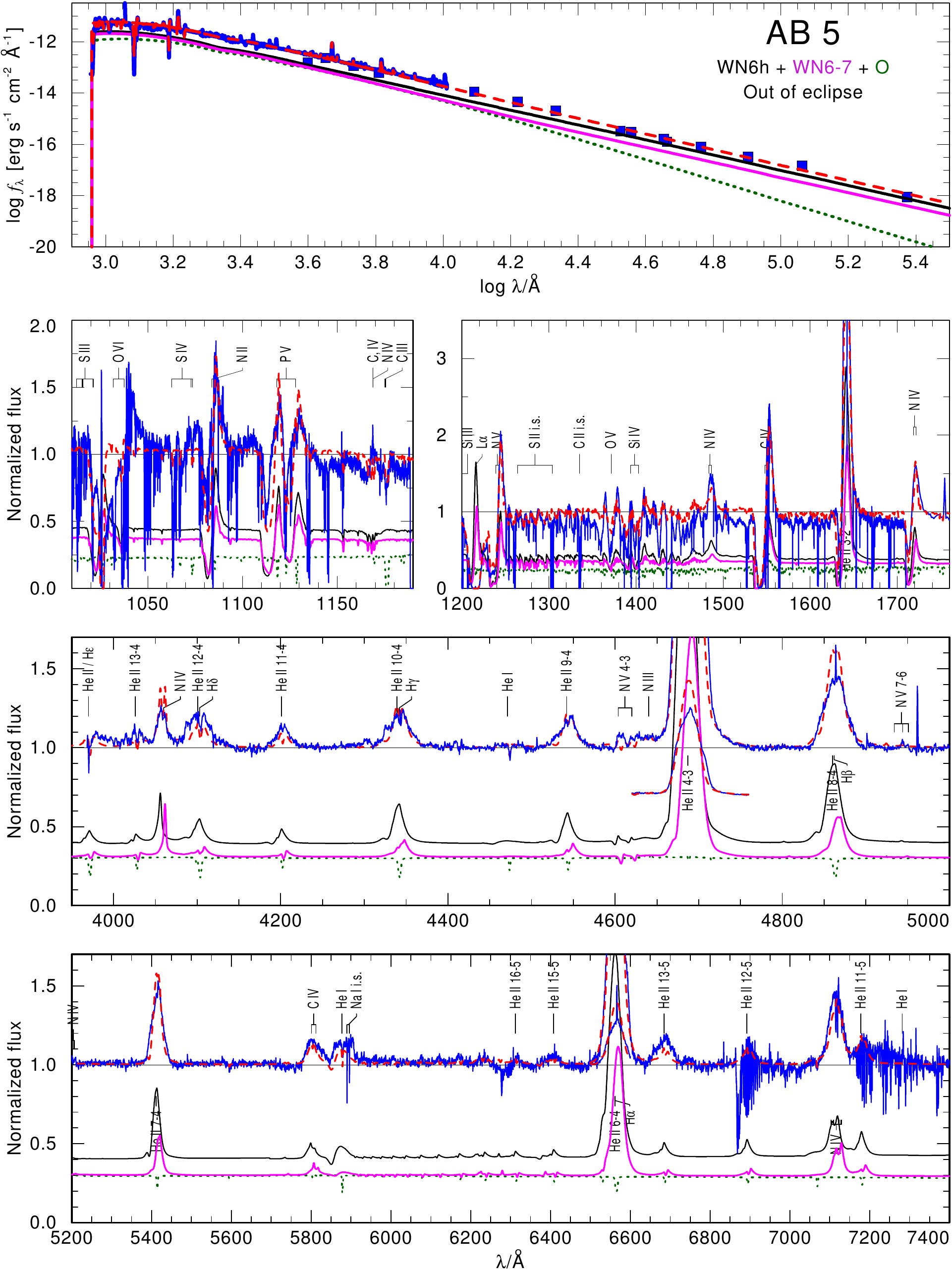}
  \caption{Same as Fig.\,\ref{fig:AB5bin}, but showing a fit to out-of-eclipse spectra of SMC\,AB\,5. The UV spectra were 
  taken with FUSE in 2002 (ID:p223010700000, $\phi = 0.74$) and HST in 1999 (ID:O55Q01070, $\phi = 0.83$). The optical 
  spectrum was taken with FEROS in 2005 (ID:f0921331, $\phi = 0.81$). }
\label{fig:AB5trip}
\end{figure*} 

\begin{figure*}[!htb]
\centering
  \includegraphics[width=\textwidth]{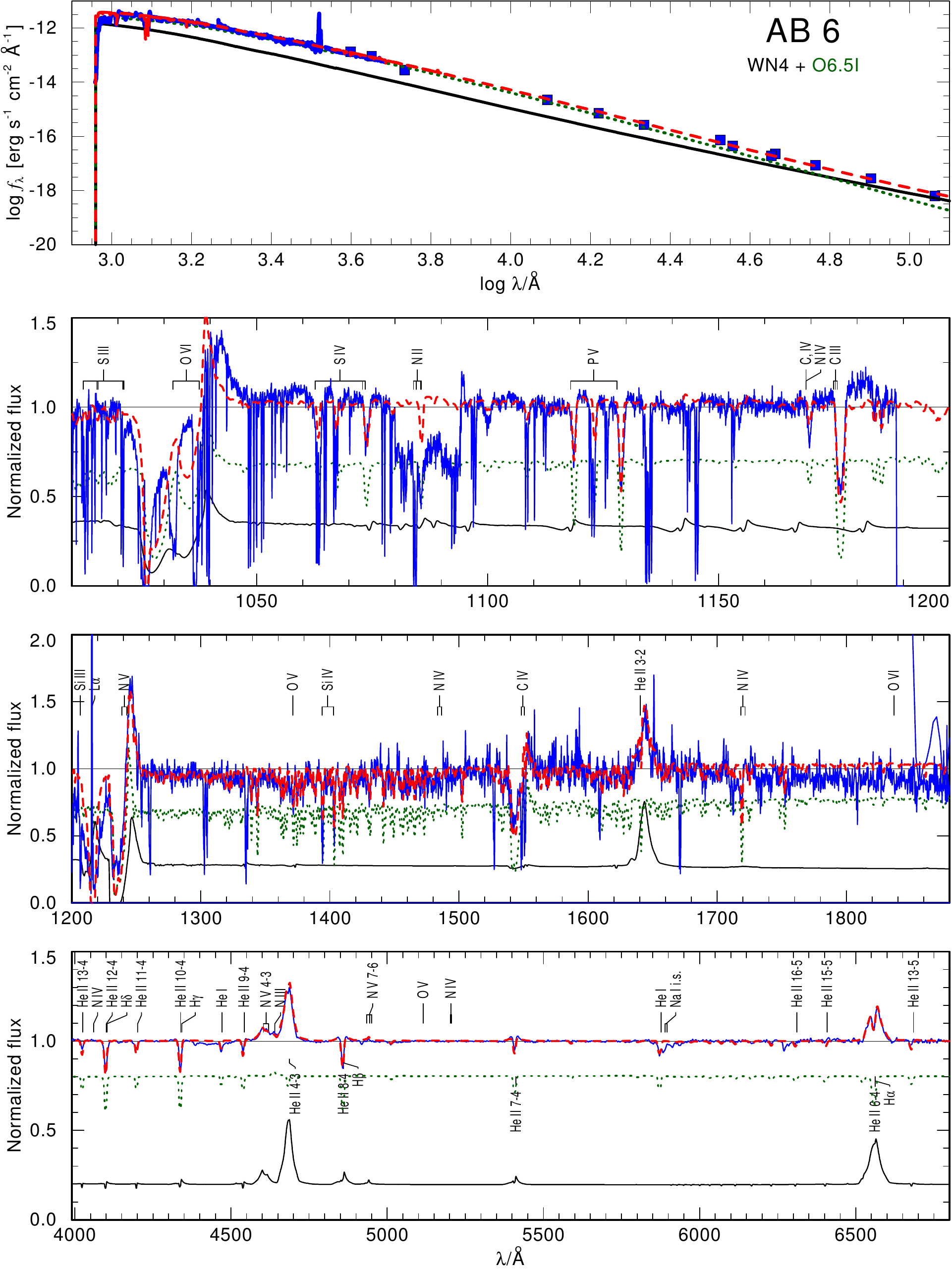}
  \caption{Same as Fig.\,\ref{fig:AB3}, but for SMC\,AB\,6. The UV spectra were taken with 
  FUSE in 2000 (ID:p103040100000, $\phi =0.42$) and IUE in 1997 (ID:sp41784, $\phi = 0.48$). The optical spectrum 
  consists of co-added spectra (see Sect.\,\ref{sec:data}).}
\label{fig:AB6}
\end{figure*} 

\begin{figure*}[!htb]
\centering
  \includegraphics[width=\textwidth]{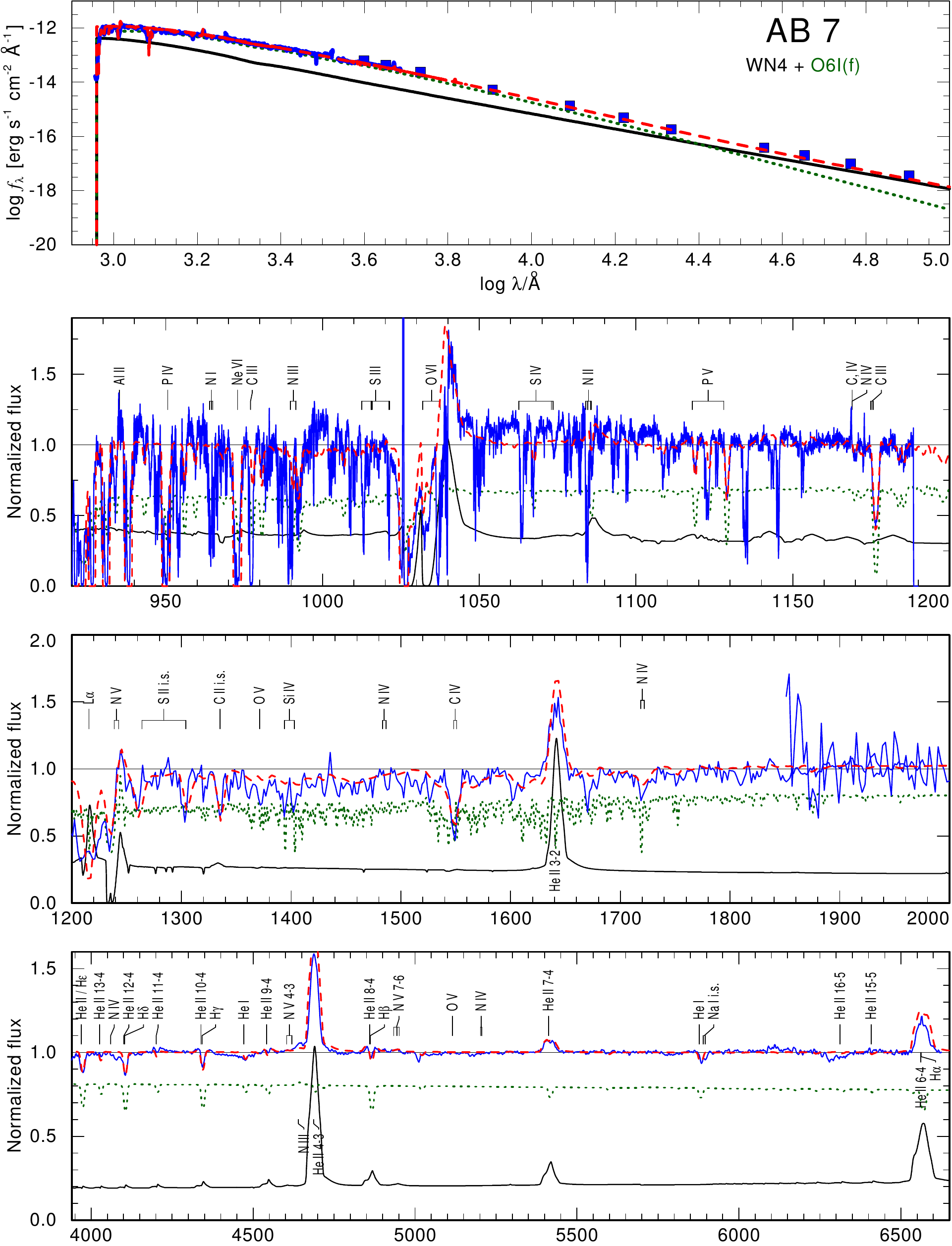}
  \caption{Same as Fig.\,\ref{fig:AB3}, but for SMC\,AB\,7. The UV spectra were taken with FUSE 
  in 2001 (IDLp2430101000, $\phi = 0.93$) and IUE in 1988 (ID:SWP33300, $\phi = 0.69$). The optical spectrum consists 
  of co-added spectra (see Sect.\,\ref{sec:data}).}
\label{fig:AB7}
\end{figure*} 

\begin{figure*}[!htb]
\centering
  \includegraphics[width=\textwidth]{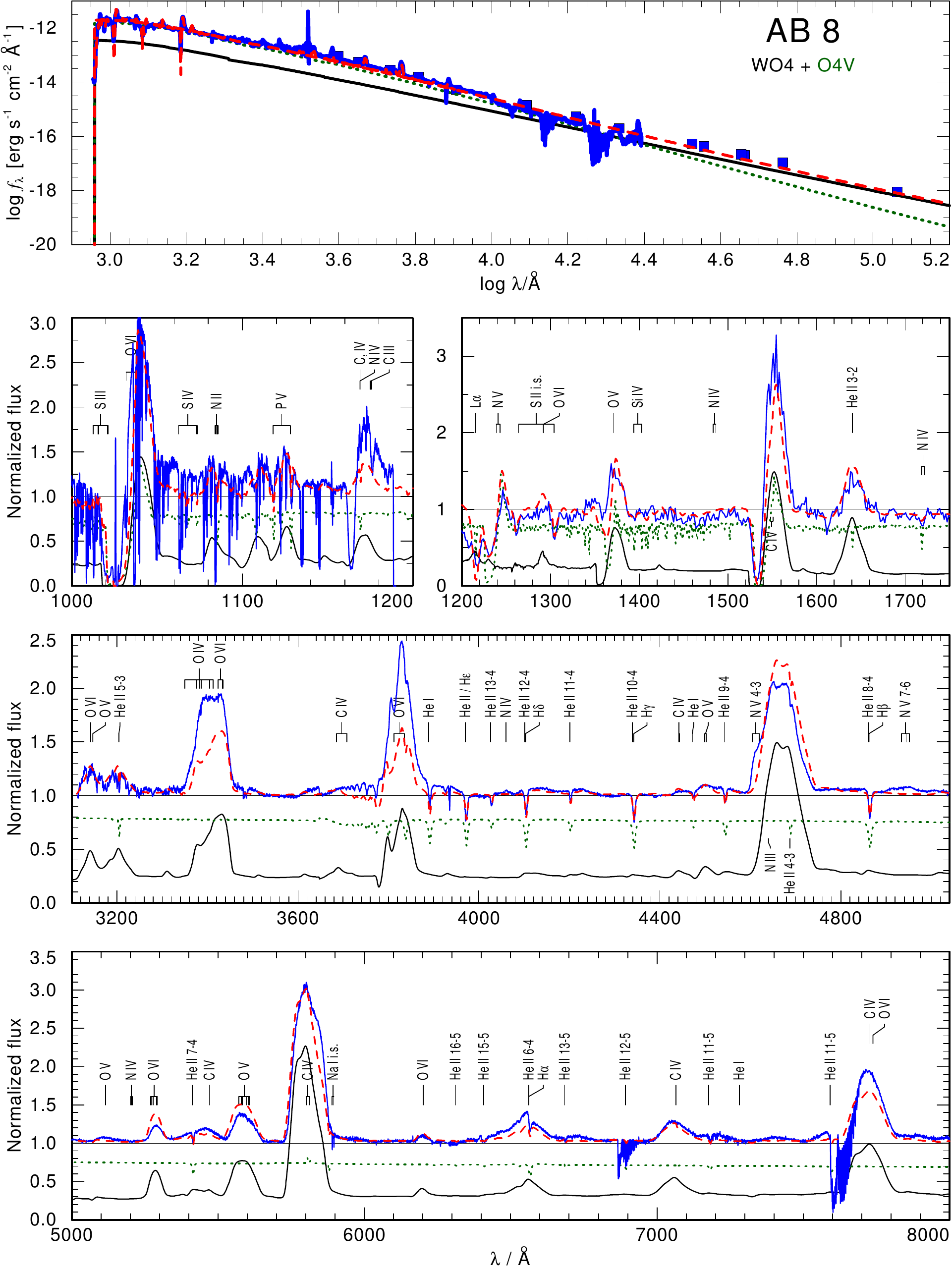}
  \caption{Same as Fig.\,\ref{fig:AB3}, but for SMC\,AB\,8. The UV spectra were taken with FUSE 
  in 2000 and IUE in 1980 (ID:sp07623, $\phi = 0.79$). The FUSE spectrum consists of co-added spectra around phase $\phi = 0.5$. 
  The optical spectrum was taken with XSHOOTER in 2013 (see Sect.\,\ref{sec:data}).}
\label{fig:AB8}
\end{figure*} 

\end{appendix} 

\end{document}